# The Relativistic Blackbody Spectrum in Inertial and Non-Inertial Reference Frames

Jeffrey S. Lee[1]
Gerald B. Cleaver[1,2]

[1]Early Universe Cosmology and Strings Group,
Center for Astrophysics, Space Physics, and Engineering Research
[2]Department of Physics
Baylor University
One Bear Place
Waco, TX 76706

Jeff_Lee@Baylor.edu
Gerald_Cleaver@Baylor.edu

## Abstract

By invoking inverse temperature as a van Kampen-Israel future-directed timelike 4-vector, this paper derives the Relativistic Blackbody Spectrum, the Relativistic Wien's Displacement Law, and the Relativistic Stefan-Boltzmann Law in inertial and non-inertial reference frames.

## 1. Introduction

The semi-empirical derivation and applications of the blackbody spectrum for exclusively stationary radiation sources have been well-established and are in many physics textbooks. However, significant progress in establishing the relativistic blackbody spectrum has been stymied, at least to some extent, by unresolved issues in relativistic thermodynamics.

This paper makes use of the inverse temperature 4-vector to derive the Relativistic Planck's Law, the Relativistic Wien's Displacement Law, and the Relativistic Stefan-Boltzmann Law in inertial reference frames. In order to describe correctly the relativistic blackbody spectrum, *relativistic beaming* and *Doppler shifting*, in addition to *relativistic temperature transformation*, must be considered. Additionally, the non-inertial reference frame case is established with the azimuthally-constant 4-acceleration and, when necessary, the proper time derivative of the spectral radiance, the wavelength of maximum irradiance, and the radiation irradiance. Also, non-trivial solutions are sought for equal spectral radiances, equal wavelengths of maximum irradiance, and equal irradiances of the relativistic and radiation frame blackbody spectra.

## 2. The Application of Inverse Temperature to the Blackbody Spectrum

Although attempts have been made to develop the Relativistic Blackbody Spectrum [1], [2], [3], these endeavors have been unsuccessful due, at least in part, to unresolved issues in relativistic thermodynamics [4], [5], [6], [7], [8], [9], [10], [11]. Disputes have arisen supporting three published Lorentz group transformations: Temperature Deflation [6], [7] and Temperature Inflation [12], [13], [14] (which can be operationally quantified with a relativistic Carnot cycle [15], [16], [17]), and Temperature Invariance [18], [19], [20], [21].

Significant misperceptions have arisen concerning temperature in relativistic thermodynamics due in part to the confusion surrounding the respective differences between empirical and absolute temperatures. The *empirical temperature* is a Lorentz invariant, relativistic scalar that considers the radiation rest frame and the observer frame to be in thermal equilibrium [22]. This ensues from the Zeroth Law of Thermodynamics, and correlates directly to the absolute temperature in the radiation (source) frame. The Zeroth Law's validity is required without making use of any thermodynamic property (including entropy and energy) [23].

The *absolute temperature* of a thermodynamic system is a consequence of the Second Law of Thermodynamics. It is the product of the Lorentz factor and the absolute temperature in the radiation frame, and contains no angular dependence. Even though the difference between empirical and absolute temperatures may be observable in non-relativistic thermodynamics, it becomes persuasively illuminated in relativistic thermodynamics.

The Planck distribution describes a solid angular photon number density, and defines a *directional* (or *effective*) temperature. However, this results from solely mathematical manipulations, and its thermodynamic relevance is, at best, questionable. Alternatively, temperature transformations can be accomplished by treating *inverse temperature* as a van Kampen-Israel future-directed timelike 4-vector. Although Przanowski and Tosiek [24][i], and Lee and Cleaver [4][ii] have demonstrated temperature inflation without making use of inverse temperature, angular dependence is required for the relativistic blackbody spectrum.

## 3. Derivation of the Relativistic Spectral Radiance

The relativistic blackbody spectrum can be obtained by considering the blackbody spectrum of a stationary radiation source, and including temperature inflation (in terms of inverse temperature), Doppler shifting, and relativistic beaming. The inertial and non-inertial frames cases are each examined. In the non-inertial case, the Unruh Effect is not considered because it is many orders of magnitude smaller than the effect presented here.

### 3.1 Inertial Frames

The radiation (source) frame photon energy density $\varepsilon$ in frequency and wavelength spaces of a Planckian distribution are:

$$\varepsilon_{\nu} d\nu = \frac{\left(\dfrac{8\pi h \nu^3}{c^3}\right)}{\exp\left(\dfrac{h\nu}{k_B T_o}\right) - 1} d\nu \tag{1}$$

and

[i] By using a superfluidity gedanken experiment.
[ii] By rewriting the stress energy tensor using occupation number.

$$\varepsilon_\lambda d\lambda = \frac{\left(\dfrac{8\pi hc}{\lambda^5}\right)}{\exp\left(\dfrac{hc}{k_B \lambda T_o}\right) - 1} d\lambda \tag{2}$$

where $h$ is Planck's constant, $c$ is the speed of light, $k_B$ is Boltzmann's constant, $T_o$ is the rest frame absolute temperature, $v$ is the frequency, and $\lambda$ is the wavelength.

Relativistically, the reciprocal of absolute temperature is replaced by the inverse temperature 4-vector[iii]:

$$\beta_\mu = \frac{u_\mu}{T_o} = \beta_t - \beta_z \cos\theta \tag{3}$$

Although the azimuthally constant $\beta_\mu$ is the reciprocal of the effective temperature, given by eq. (4), the inverse temperature arises directly from thermodynamic considerations, whereas the effective temperature is obtained entirely from mathematical manipulation. Although $T_{\text{eff}}$ is adequate to determine planetary and stellar motion with respect to the CMB [25], [26], [27], its thermodynamically non-physical origin leaves it unclear whether or not $T_{\text{eff}}$ represents temperature.

$$T_{\text{eff}} = \frac{T_o \sqrt{1 - V^2}}{1 - V\cos\theta} \tag{4}$$

Rewriting eqs. (1) and (2) in terms of inverse temperature yields[iv]:

$$\varepsilon_\nu^{'} d\nu d\Omega = \frac{\left(\dfrac{8\pi h\nu^3}{c^3}\right)}{\exp\left[\dfrac{h\nu}{k_B}\left(\beta_t - \beta_z \cos\theta\right)\right] - 1} d\nu d\Omega \tag{5}$$

and

[iii] In this paper, the case of the azimuthally constant inverse temperature vector is chosen (hence, $\beta_x = \beta_y = 0$), and consequently, all motion is along the $z$-axis.

[iv] Primed quantities indicate the observer frame.

$$\varepsilon'_{\lambda} d\lambda d\Omega = \frac{\left(\frac{8\pi hc}{\lambda^5}\right)}{\exp\left[\frac{hc}{k_B \lambda}\left(\beta_t - \beta_z \cos\theta\right)\right] - 1} d\lambda d\Omega \tag{6}$$

where [28]:

$$\beta_t = \frac{1}{T_o \sqrt{1 - V^2}} \tag{7}$$

$$\beta_z = \frac{V}{T_o \sqrt{1 - V^2}} \tag{8}$$

$T_o$ is the proper temperature (in the radiation frame).

$u_\mu$ is the relative 4-velocity between the radiation and the observer.

$\beta_\mu$ is the van Kampen-Israel inverse temperature 4-vector.

$V = \frac{u}{c}$ (fraction of light speed).

From $B'_{\nu,\lambda} = \frac{c}{4\pi}\varepsilon'_{\nu,\lambda}$, $\frac{B'_\lambda}{B_\lambda} = \left[\gamma\left(1 - V\cos\theta\right)\right]^{-3} = D^3$ (where $\gamma = \frac{1}{\sqrt{1 - V^2}}$ is the Lorentz factor, and $D$ is the Doppler factor), eqs. (5) and (6) become respectively:

$$B'_\nu d\nu d\Omega = \frac{\left(\frac{2h\nu^3}{c^2}\right)}{\exp\left[\frac{h\nu}{k_B}\left(\beta_t - \beta_z \cos\theta\right)\right] - 1}\left[\gamma\left(1 - V\cos\theta\right)\right]^{-3} d\nu d\Omega \tag{9}$$

and

$$B'_\lambda d\lambda d\Omega = \frac{\left(\frac{2hc^2}{\lambda^5}\right)}{\exp\left[\frac{hc}{k_B \lambda}\left(\beta_t - \beta_z \cos\theta\right)\right] - 1}\left[\gamma\left(1 - V\cos\theta\right)\right]^{-3} d\lambda d\Omega \tag{10}$$

where $B'_\nu$ and $B'_\lambda$ are respectively the frequency and wavelength spectral radiances in the observer frame.

Rewritten in terms of the temperature in the radiation frame, eqs. (9) and (10) become:

$$B_{\nu}^{'} d\nu d\Omega = \frac{\left(\frac{2h\nu^3}{c^2}\right)\left(1-V^2\right)^{\frac{3}{2}}}{\left(1-V\cos\theta\right)^3\left\{\exp\left[\left(\frac{h\nu}{k_B T_o}\right)\frac{\left(1-V\cos\theta\right)}{\left(1-V^2\right)^{\frac{1}{2}}}\right]-1\right\}} d\nu d\Omega \tag{11}$$

and

$$B_{\lambda}^{'} d\lambda d\Omega = \frac{\left(\frac{2hc^2}{\lambda^5}\right)\left(1-V^2\right)^{\frac{3}{2}}}{\left(1-V\cos\theta\right)^3\left\{\exp\left[\left(\frac{hc}{k_B \lambda T_o}\right)\frac{\left(1-V\cos\theta\right)}{\left(1-V^2\right)^{\frac{1}{2}}}\right]-1\right\}} d\lambda d\Omega \tag{12}$$

The dependency on wavelength of the spectral radiance in wavelength space is shown in Figure 1.

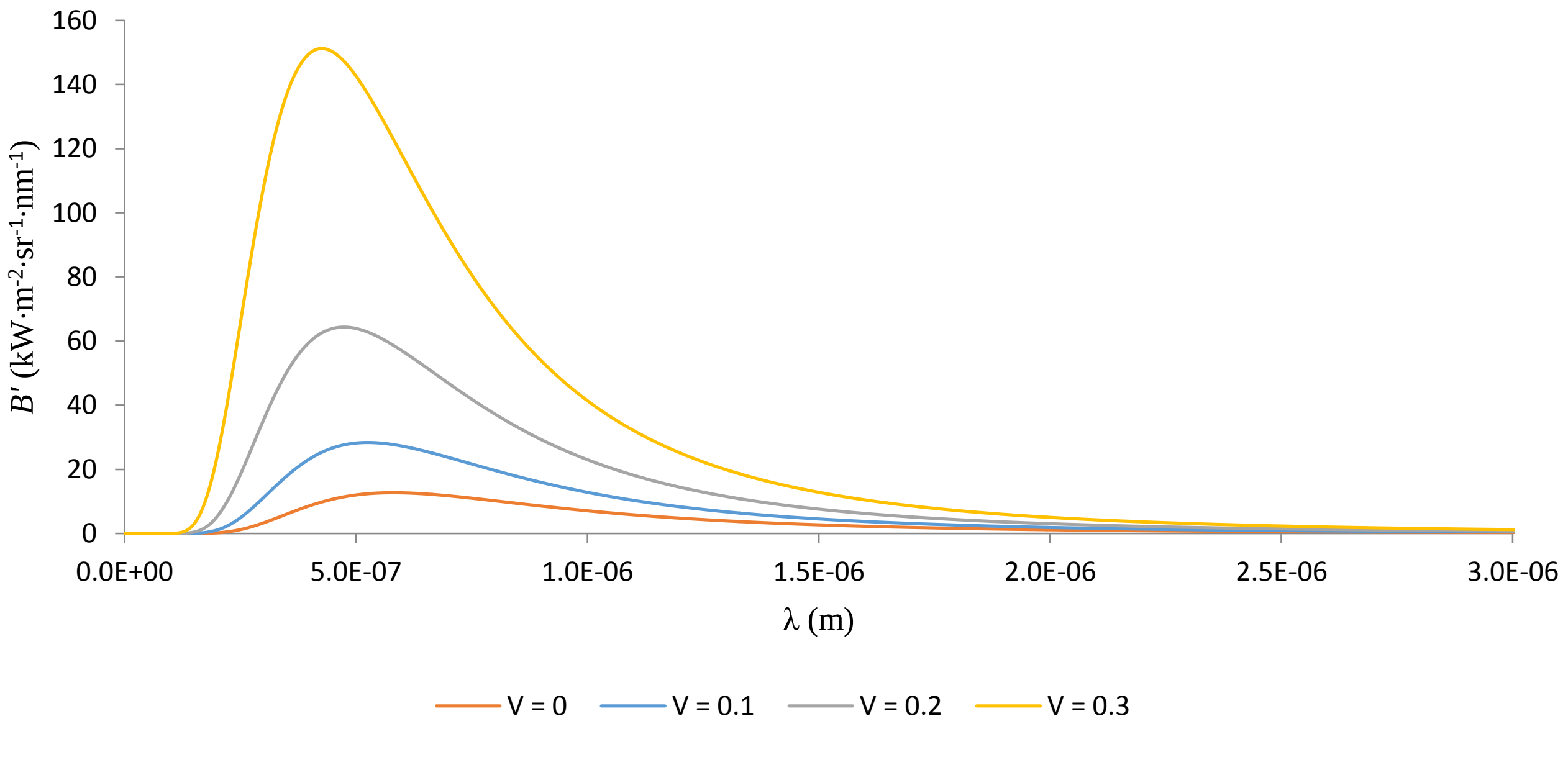

Figure 1: Spectral radiance versus wavelength for a 5000 K radiation source in four inertial reference frames. $\theta = 0$.

The dependency on speed of the spectral radiance in wavelength space is shown in Figure 2.

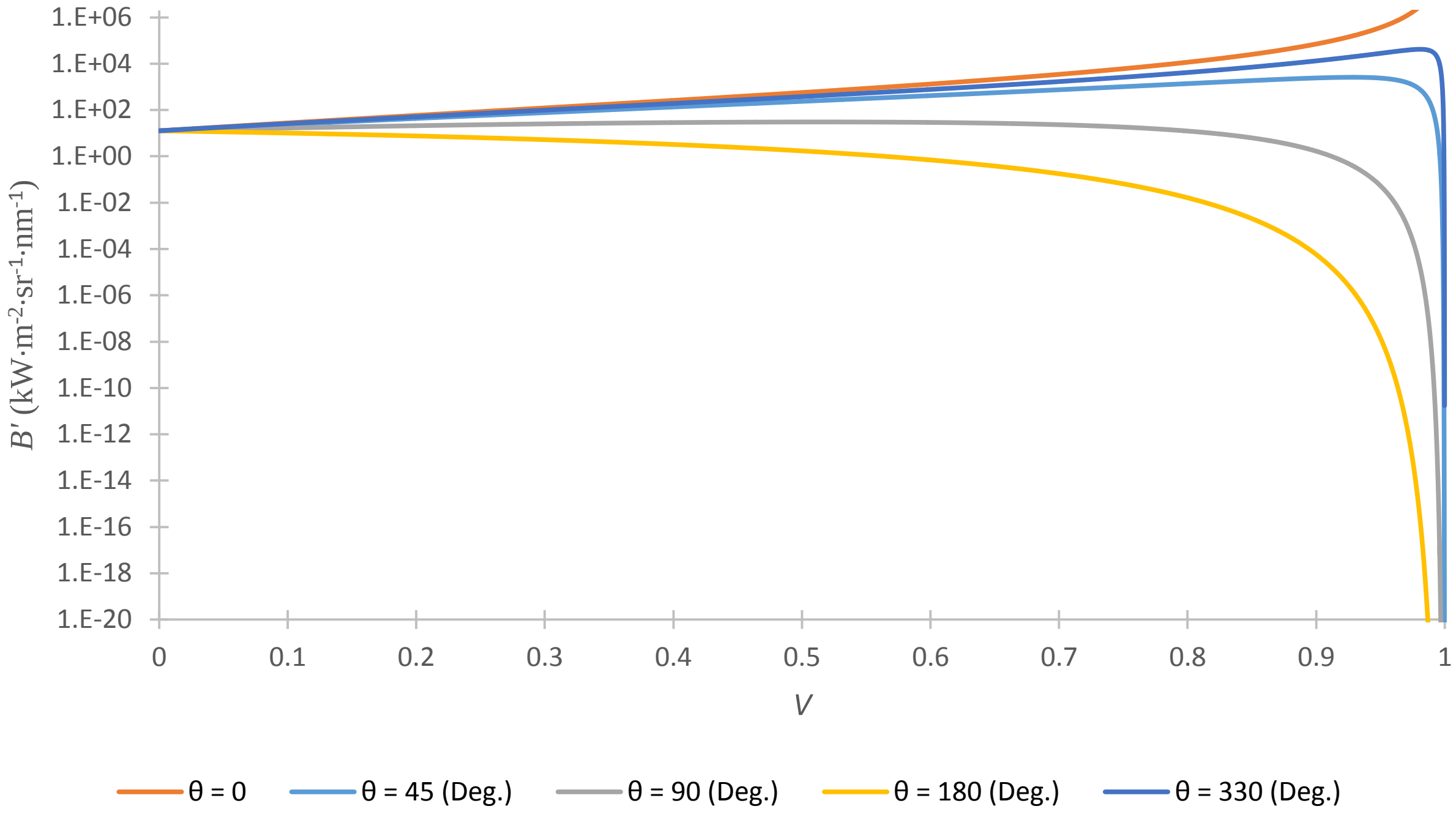

Figure 2: Spectral radiance versus speed at five angles for a 5000 K blackbody at the wavelength of maximum irradiance in the radiation frame (0.5796 $\mu$m). The $B'$-axis intercept is the spectral radiance in the radiation frame (12.7 kW·m$^{-2}$·sr$^{-1}$·nm$^{-1}$).

The contour lines on the $V$-$\theta$ contour plot (Figure 3) of eq. (12) reveal that $B'$ increases with increasing $V$ only for first and fourth quadrant angles ( $\cos\theta > 0$ ), while it decreases for second and third quadrant angles ( $\cos\theta < 0$ ). Expectedly, the inertial relativistic spectral radiance decreases most appreciably with increasing speed when $\theta = 180^{\circ}$. Thus, the 4-vector velocity behavior of relativistic spectral radiance does not trivially increase $B'$ for increasing $\theta$.

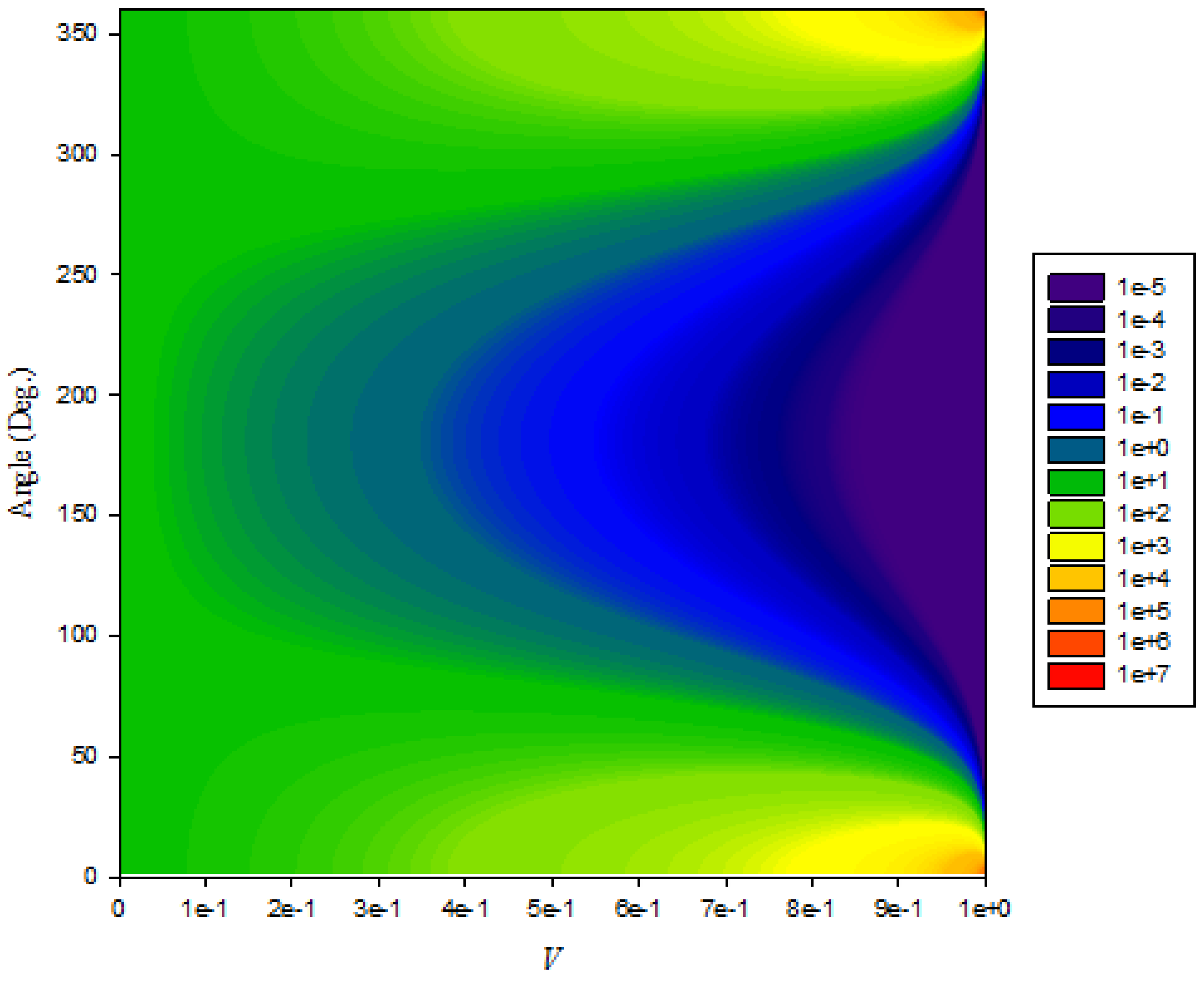

Figure 3: Contour plot of inertial relativistic spectral radiance as a function of speed and angle for a 5000 K blackbody at the wavelength of maximum irradiance in the rest frame (0.5796 $\mu$m). The color spectrum gives the approximate values of $B'$ ($\mathrm{kW \cdot m^{-2} \cdot sr^{-1} \cdot nm^{-1}}$). The contour lines' intersections with the *Angle*-axis represent the spectral radiance in the rest frame, which is 12.7 $\mathrm{kW \cdot m^{-2} \cdot sr^{-1} \cdot nm^{-1}}$. The region of smallest spectral radiance is actually $B' \leq 10^{-5}$ $\mathrm{kW \cdot m^{-2} \cdot sr^{-1} \cdot nm^{-1}}$.

Several authors [2], [3], [29] have remarked that the Bose-Einstein distribution form of the blackbody spectrum of a stationary radiation source is not relativistically invariant. However, eqs. (11) and (12) clearly reveal that, in terms of the radiation frame temperature, the Bose-Einstein distribution for spectral radiance is retained in a relativistic inertial reference frame (as shown in Figure 1). Regardless, inverse temperature remains a valid thermodynamic quantity [2].

In the non-relativistic limit, eq. (12) becomes:

$$B_\lambda' d\lambda d\Omega \sim \left\{ \frac{\left(\frac{2hc^2}{\lambda^5}\right)}{\exp\left(\frac{hc}{k_B \lambda T_o}\right)-1} + \left[ \frac{\left(\frac{6hc^2}{\lambda^5}\right)}{\exp\left(\frac{hc}{k_B \lambda T_o}\right)-1} + \frac{\left(\frac{2h^2c^3}{\lambda^6}\right)\exp\left(\frac{hc}{k_B \lambda T_o}\right)\cos\theta}{\left[\exp\left(\frac{hc}{k_B \lambda T_o}\right)-1\right]^2} \right] V + O\left(V^2\right) \right\} d\lambda d\Omega \qquad (13)$$

From eq. (12), $V < 1$ clearly disallows any non-zero speed with respect to the radiation frame for which $B' = 0$, and also, since $\cos\theta \le 1$, there is no subluminal speed for which $B'$ is infinite. When $V = 0$, then $\beta_z = 0$, $\beta_t = \frac{1}{T_o}$ (the zeroth component of inverse temperature in the radiation frame), and $\gamma = 1$. Therefore, as required, the stationary forms of eqs. (11) and (12) are recovered. This result is, of course, clear from eq. (13) as well.

$$B_\nu = \frac{\left(\frac{2h\nu^3}{c^2}\right)}{\exp\left[\frac{h\nu}{k_B T_o}\right] - 1} \tag{14}$$

and

$$B_\lambda = \frac{\left(\frac{2hc^2}{\lambda^5}\right)}{\exp\left[\frac{hc}{k_B \lambda T_o}\right] - 1} \tag{15}$$

The inertial relativistic blackbody spectra for other spaces are summarized in Table 1.

| Variable | Distribution |
| --- | --- |
| Frequency $(\nu)$ | $B_{\nu}^{'} d\nu d\Omega = \dfrac{\left(\dfrac{2h\nu^{3}}{c^{2}}\right)\left(1-V^{2}\right)^{\frac{3}{2}}}{\left(1-V\cos\theta\right)^{3}\left\{\exp\left[\left(\dfrac{h\nu}{k_{B}T_{o}}\right)\dfrac{\left(1-V\cos\theta\right)}{\left(1-V^{2}\right)^{\frac{1}{2}}}\right]-1\right\}} d\nu d\Omega$ |
| Wavelength $(\lambda)$ | $B_{\lambda}^{'} d\lambda d\Omega = \dfrac{\left(\dfrac{2hc^{2}}{\lambda^{5}}\right)\left(1-V^{2}\right)^{\frac{3}{2}}}{\left(1-V\cos\theta\right)^{3}\left\{\exp\left[\left(\dfrac{h\nu}{k_{B}T_{o}}\right)\dfrac{\left(1-V\cos\theta\right)}{\left(1-V^{2}\right)^{\frac{1}{2}}}\right]-1\right\}} d\lambda d\Omega$ |
| Wavenumber $(\tilde{\nu})$ | $B_{\tilde{\nu}}^{'} d\tilde{\nu} d\Omega = \dfrac{2hc^{2}\tilde{\nu}^{3}}{\left(1-V^{2}\right)^{\frac{3}{2}}\left(1-V\cos\theta\right)^{3}\left\{\exp\left[\left(\dfrac{h\nu}{k_{B}T_{o}}\right)\dfrac{\left(1-V\cos\theta\right)}{\left(1-V^{2}\right)^{\frac{1}{2}}}\right]-1\right\}} d\tilde{\nu} d\Omega$ |
| Angular Frequency $(\omega)$ | $B_{\omega}^{'} d\omega d\Omega = \dfrac{\left(\dfrac{\hbar\omega^{3}}{4\pi^{3}c^{2}}\right)\left(1-V^{2}\right)^{\frac{3}{2}}}{\left(1-V\cos\theta\right)^{3}\left\{\exp\left[\left(\dfrac{h\nu}{k_{B}T_{o}}\right)\dfrac{\left(1-V\cos\theta\right)}{\left(1-V^{2}\right)^{\frac{1}{2}}}\right]-1\right\}} d\omega d\Omega$ |
| Angular Wavelength $(y)$ | $B_{y}^{'} d\lambda d\Omega = \dfrac{\left(\dfrac{\hbar c^{2}}{4\pi^{3}y^{5}}\right)\left(1-V^{2}\right)^{\frac{3}{2}}}{\left(1-V\cos\theta\right)^{3}\left\{\exp\left[\left(\dfrac{h\nu}{k_{B}T_{o}}\right)\dfrac{\left(1-V\cos\theta\right)}{\left(1-V^{2}\right)^{\frac{1}{2}}}\right]-1\right\}} dy d\Omega$ |
| Angular Wavenumber $(k)$ | $B_{k}^{'} d\omega d\Omega = \dfrac{\left(\dfrac{\hbar c^{2}k^{3}}{4\pi^{3}}\right)\left(1-V^{2}\right)^{\frac{3}{2}}}{\left(1-V\cos\theta\right)^{3}\left\{\exp\left[\left(\dfrac{h\nu}{k_{B}T_{o}}\right)\dfrac{\left(1-V\cos\theta\right)}{\left(1-V^{2}\right)^{\frac{1}{2}}}\right]-1\right\}} dk d\Omega$ |

Table 1: Inertial relativistic blackbody spectra in terms of six spectral variables.

### 3.1.1 The Angular Periodicity of the Inertial Relativistic Spectral Radiance

Due to $\cos\theta$ in eq. (12), the relativistic spectral radiance exhibits an angular periodicity with an angular wavelength of 360°. Fifty-eight local maxima occur resulting in an "embedded" angular wavelength of 360°/57~6.3216°. This result is independent of speed (Figure 4) and temperature when the radiation frame wavelength of maximum irradiance is considered (Figure 5); this is, of course, reasonable because the temperature dependence in the angular term of eq. (12) vanishes as a result of Wien's Displacement Law applied in the rest frame, as shown in eq. (16).

$$\exp\left[\left(\frac{hc}{k_B\lambda\cdot\left(T_0=\frac{b}{\lambda}\right)}\right)\frac{(1-V\cos\theta)}{(1-V^2)^{\frac{1}{2}}}\right]=\exp\left[\left(\frac{hc}{k_B b}\right)\frac{(1-V\cos\theta)}{(1-V^2)^{\frac{1}{2}}}\right] \tag{16}$$

where $b \sim 2.8977721\times10^6$ nm·K is Wien's displacement constant.

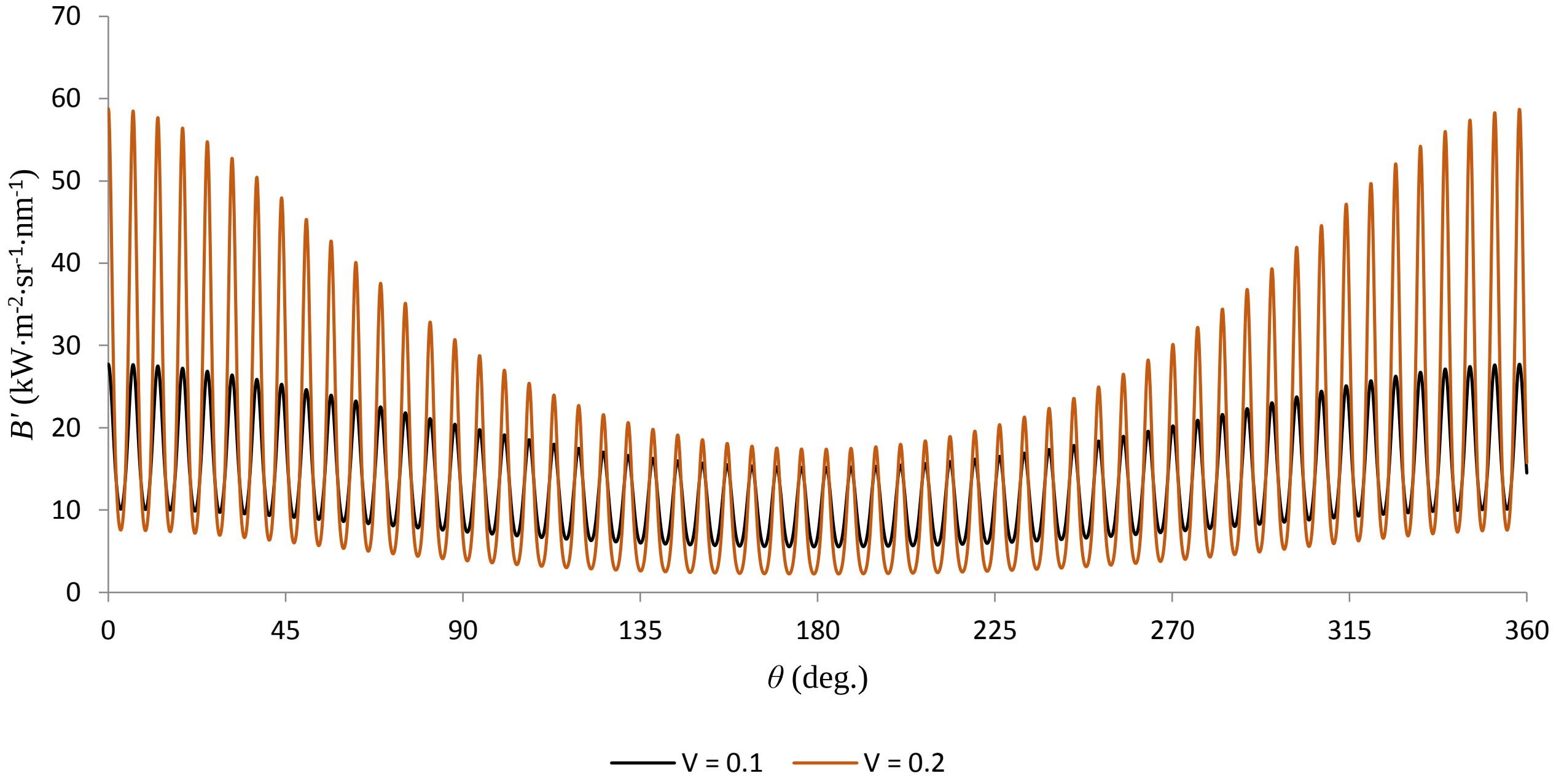

Figure 4: Inertial relativistic spectral radiance versus angle for a 5000 K blackbody at the wavelength of maximum irradiance in the rest frame (0.5796 $\mu$m).

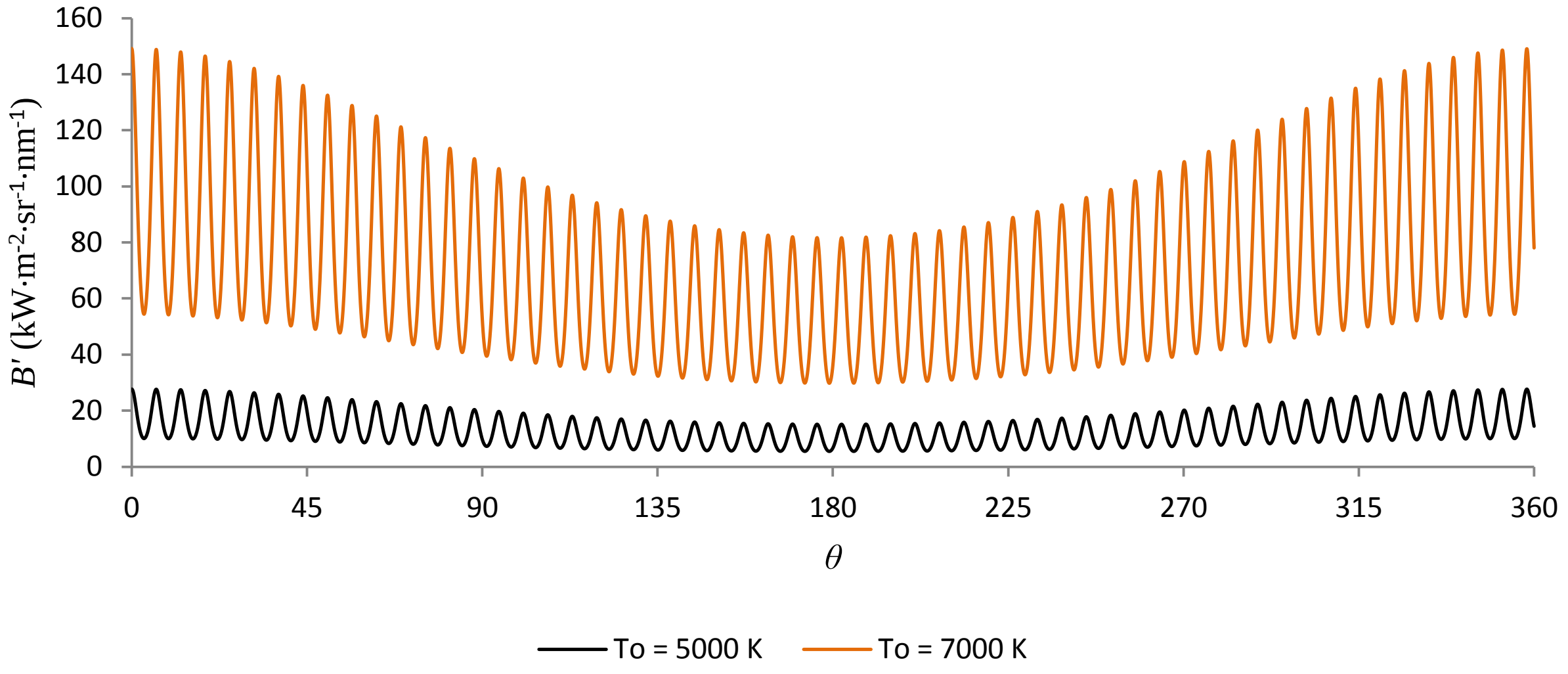

Figure 5: Relativistic spectral radiance versus angle for two blackbodies. The wavelengths are for maximum irradiance in the rest frame (0.5796 $\mu$m for 5000 K and 0.4140 $\mu$m for 7000 K). $V = 0.1$.

Although the angles at which the local maxima occur are independent of the wavelength, the local minima angles are right-shifted for increasing wavelength, and are dependent on the $\lambda T_o$ product, resulting in effects on the periodicity which are evident at two decimal places in angular degrees (four decimal places in radians) (see Figure 6 and Figure 7).

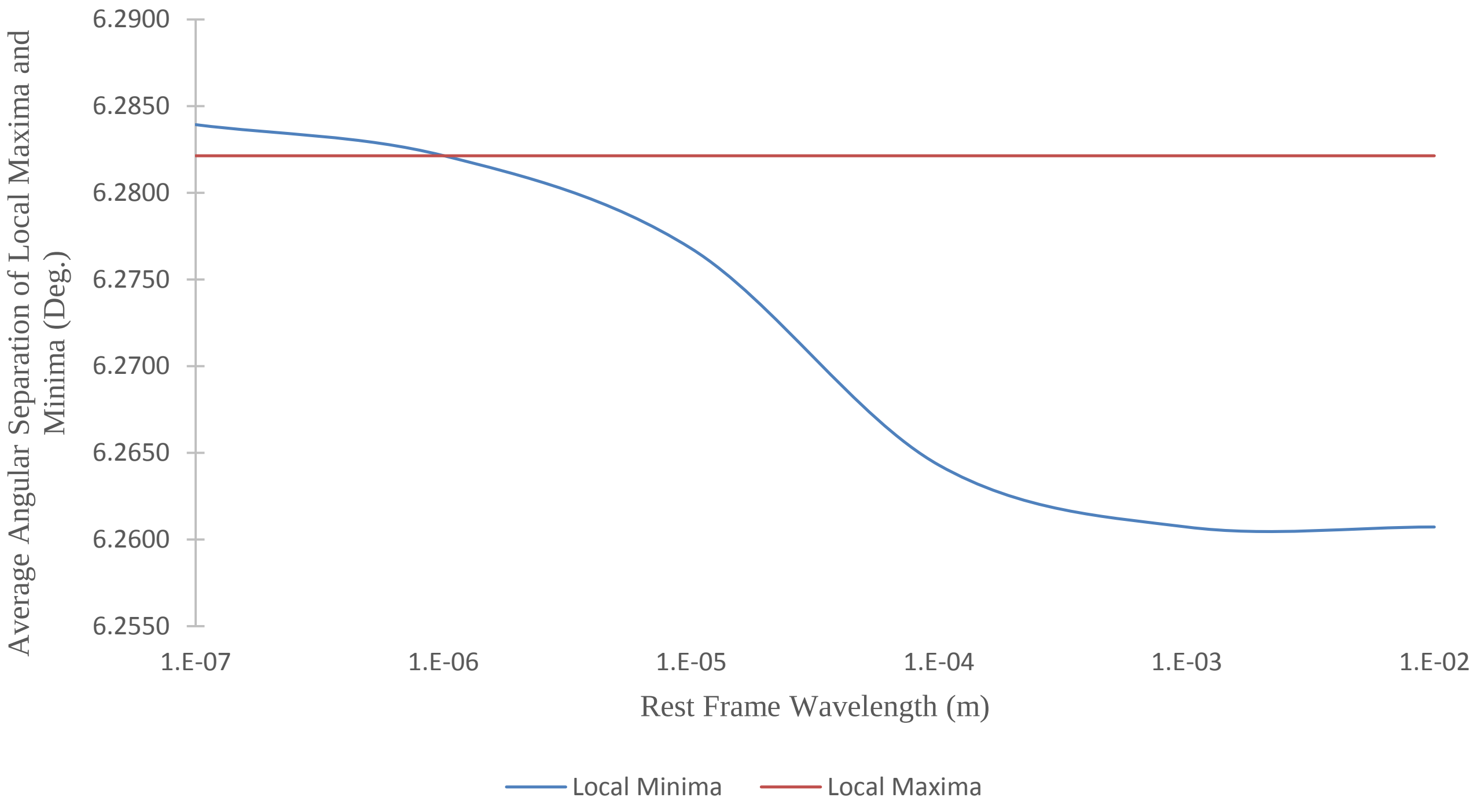

Figure 6: Average separation of local minima and maxima versus wavelength for $B'(\theta)$ of a 5000 K inertial relativistic blackbody. $V$ = 0.99.

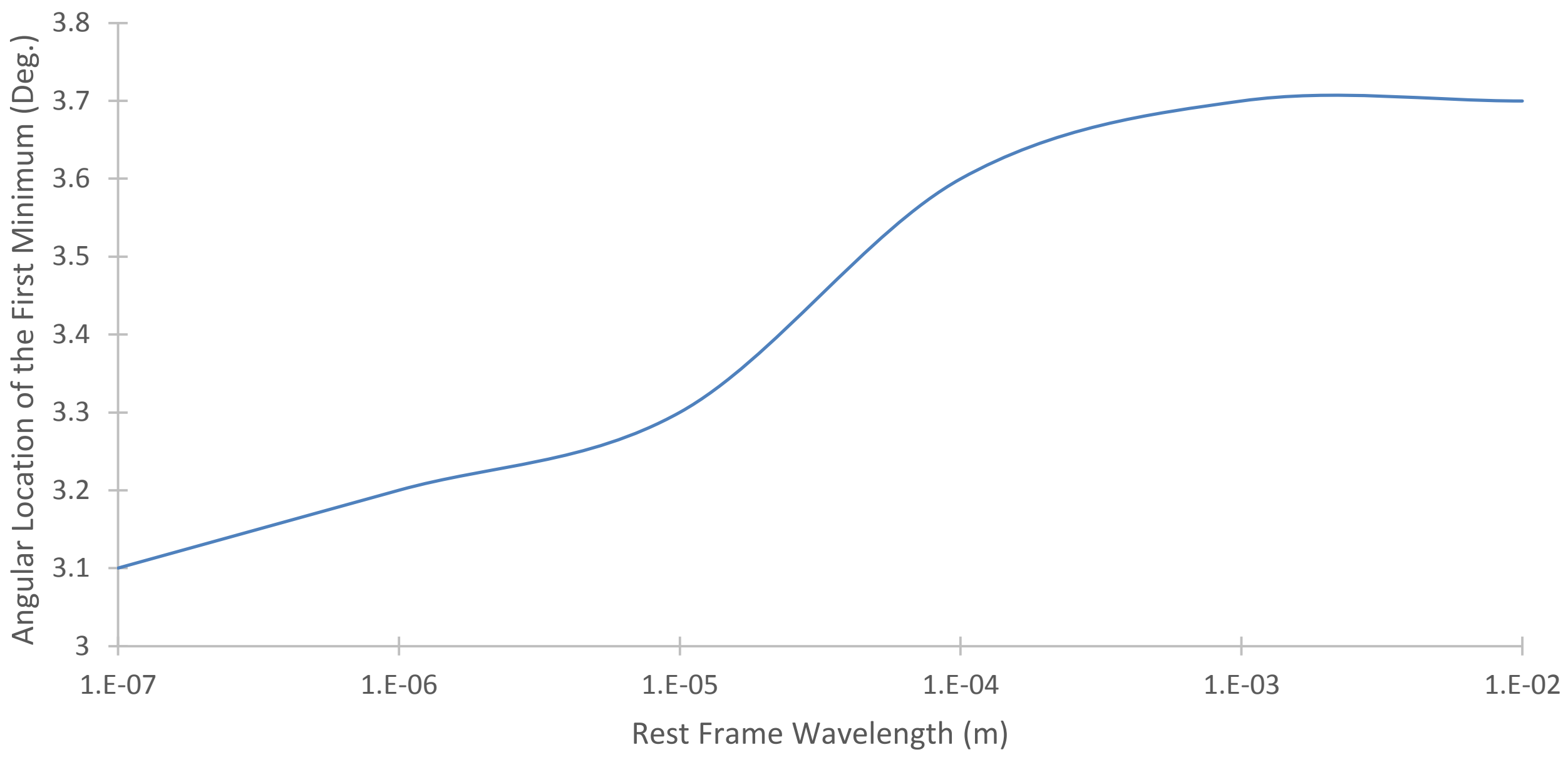

Figure 7: Location of the first local minimum versus wavelength for $B'(\theta)$ of a 5000 K inertial relativistic blackbody. $V = 0.99$.

### 3.2 Non-Inertial Reference Frames

The relativistic spectral radiance in a non-inertial reference frame is determined with the 4-acceleration $a_\mu$, which is the proper time ($\tau$) derivative (denoted with dot notation) of the 4-velocity $\left(a_\mu = \dot{u}_\mu\right)$. Combining eqs. (12) and (17), and defining $A = \frac{a}{c}$ (the zeroth term of the acceleration 4-vector), the relativistic spectral radiance in a non-inertial frame is given by eq. (18).

$$V = \tanh\left(A\tau\right) \tag{17}$$

$$B_\lambda^{'} d\lambda d\Omega = \frac{\left(\frac{2hc^2}{\lambda^5}\right)\operatorname{sech}^3\left(A\tau\right)}{\left[1 - \tanh\left(A\tau\right)\cos\theta\right]^3 \left[\exp\left(\frac{hc}{k_B \lambda T_o}\left(\cosh\left(A\tau\right) - \sinh\left(A\tau\right)\cos\theta\right)\right) - 1\right]} d\lambda d\Omega \tag{18}$$

As $A\tau \to \infty$, $B_\lambda^{'} \to \infty$ when $\theta = 0$, and $B_\lambda^{'} \to 0$ when $\theta = 90^o$ or $180^o$. As required, when $A\tau = 0$, the stationary form of the spectral radiance, eq. (15), is recovered. A plot of eq. (18) is shown in Figure 8 and Figure 9.

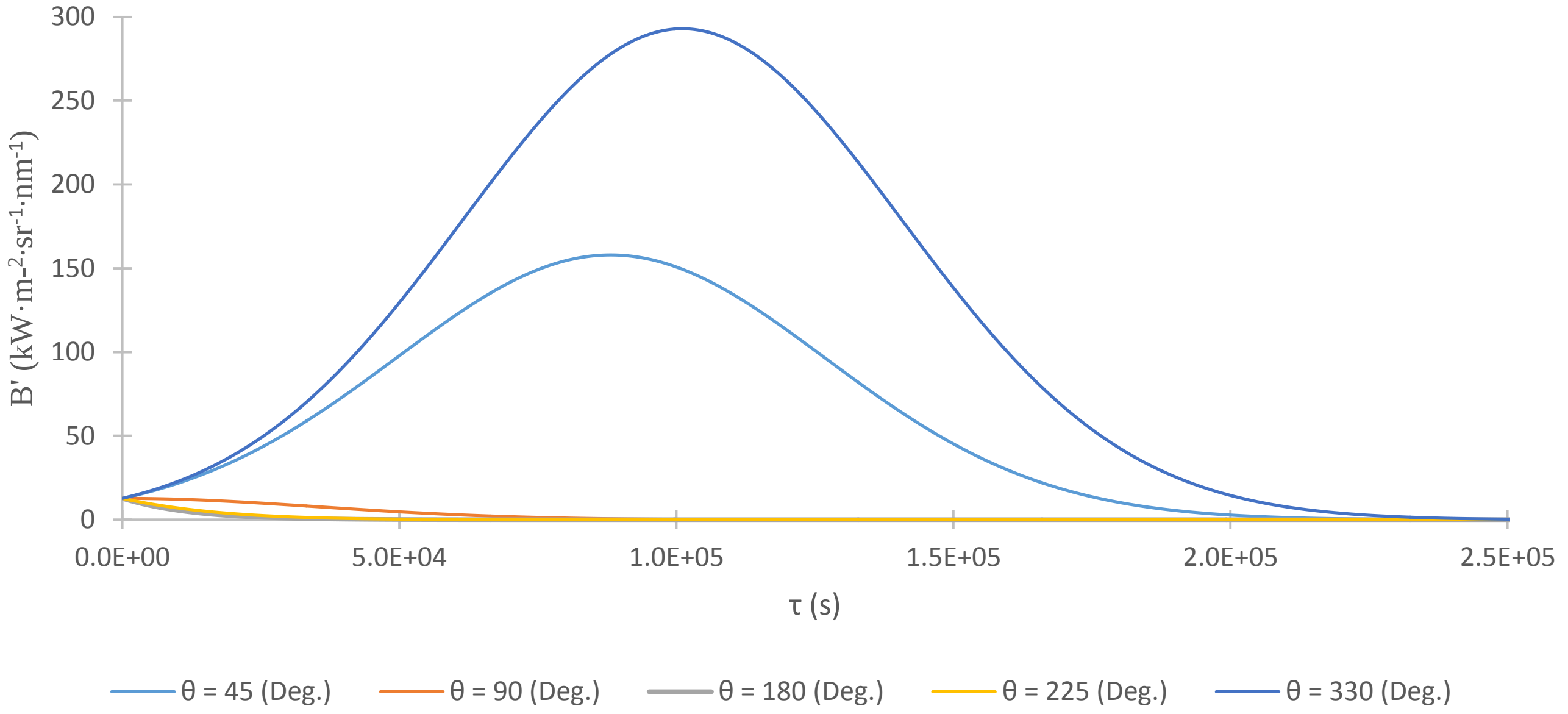

Figure 8: Non-inertial relativistic spectral radiance versus proper time for a 5000 K blackbody. $A = 10^{-5}$ $\left(a = 3\times10^{3}\ \text{m/s}^{2}\right)$. The lower left region is enlarged in Figure 9.

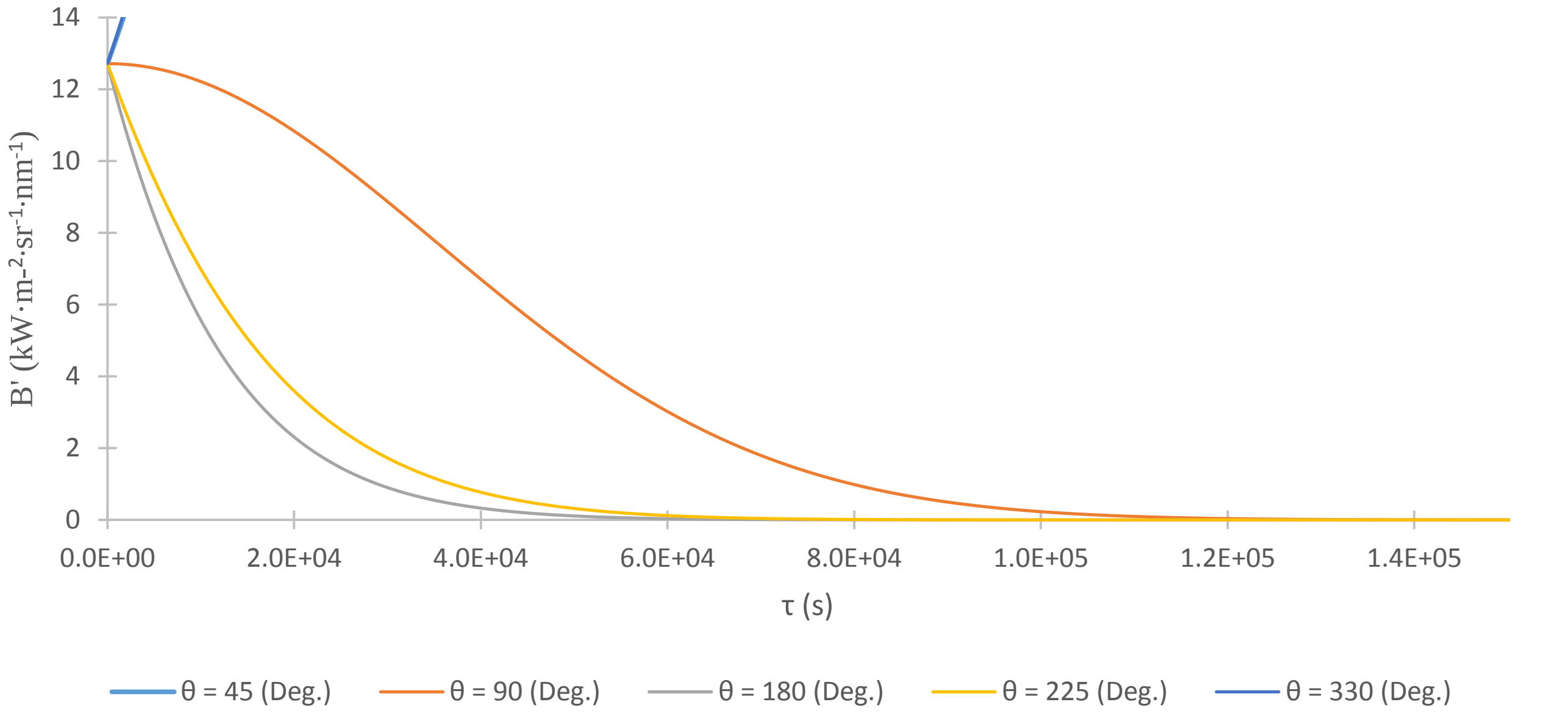

Figure 9: Non-inertial relativistic spectral radiance versus proper time for a 5000 K blackbody. $A = 10^{-5}$ $\left(a = 3\times10^{3}\ \text{m/s}^{2}\right)$.

For constant acceleration (in both magnitude and direction), the functional form of eq. (12) is unchanged. Hence, the angular dependence of the non-inertial spectral radiance is unchanged from the inertial case. However, when the proper time at which the maximum non-inertial relativistic spectral radiance occurs ($\tau_{max}$) is considered as a function of angle, Figure 10 results.

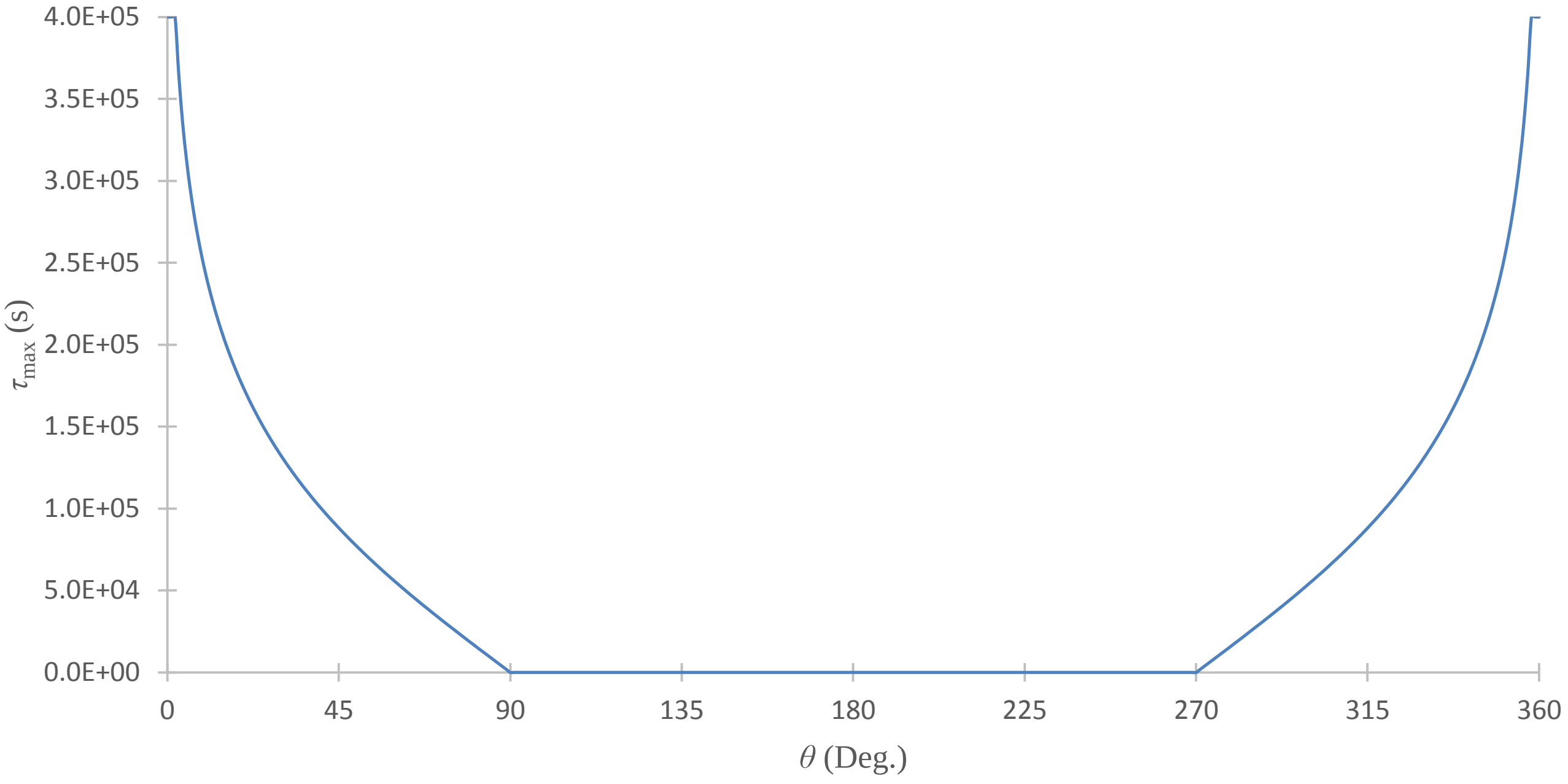

Figure 10: Time of occurrence versus angle for local maxima of the non-inertial relativistic spectral radiance. $T$ = 5000 K. $\lambda_{max}$ = 0.5796 $\mu$m. $A = 10^{-5}$.

As $A\tau \to \infty$ for $0 \le \theta \le 90^o$ or $270^o \le \theta \le 360^o$ (approaching the radiation source), the non-inertial relativistic spectral radiance becomes infinite. However, for $90^o \le \theta \le 270^o$ (receding from the radiation source), $B' \to 0$. This angular dependence is the same for the inertial relativistic spectral radiance in eq. (12). Also, as required, when $A\tau = 0$, the stationary form of the non-inertial relativistic spectral radiance, eq. (15), is recovered. This behavior was discussed briefly in Section 3.1.

The time rate of change of the relativistic spectral radiance is its proper time derivative, $\dot{B}' \equiv \frac{dB'}{d\tau}$, and is non-trivially given by eq. (19) and plotted in Figure 11.

$$\dot{B}_{\lambda}' d\lambda d\Omega = \frac{2hc^2 \operatorname{sech}^2(A\tau)}{\lambda^6 k_B T_o \left( \exp\left( \frac{hc}{k_B \lambda T_o} (\cosh(A\tau) - \sinh(A\tau)\cos\theta) \right) - 1 \right)^2 (\tanh(A\tau)\cos\theta - 1)^4} \times$$
$$\left\{ \begin{array}{l} hc \exp\left( \frac{hc}{k_B \lambda T_o} (\cosh(A\tau) - \sinh(A\tau)\cos\theta) \right) (\tanh(A\tau)\cos\theta - 1) - 3k_B \lambda T_o \operatorname{sech}(A\tau) \times \\ \left( \exp\left( \frac{hc}{k_B \lambda T_o} (\cosh(A\tau) - \sinh(A\tau)\cos\theta) \right) - 1 \right) \end{array} \right\} \times \tag{19}$$
$$\left[ \tanh(A\tau)(A + \dot{\theta}\sin\theta) - A\cos\theta \right] d\lambda d\Omega$$

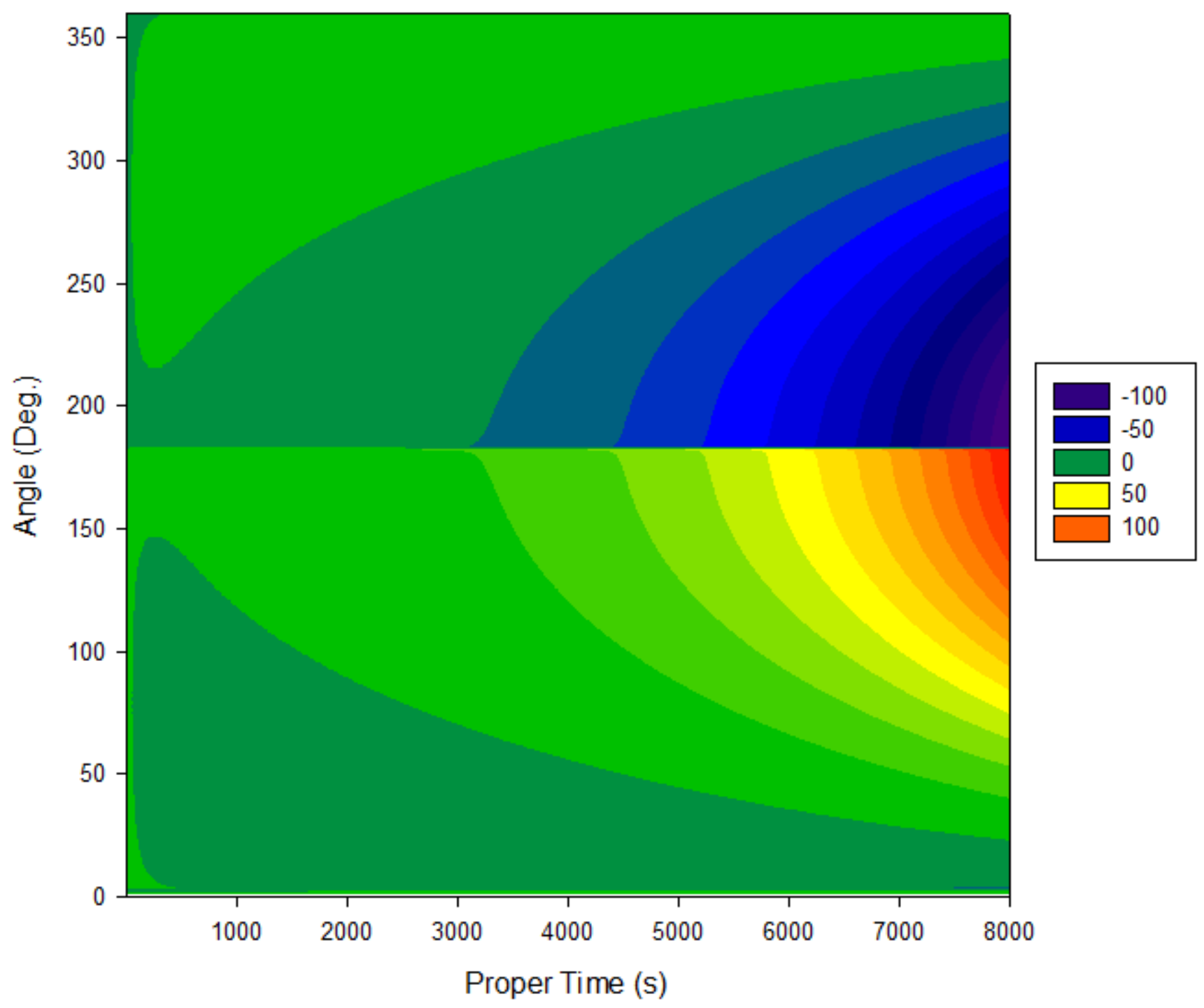

Figure 11: Contour plot of the logarithm of the time rate of change of spectral radiance as a function of angle and proper time. Negative values of $\dot{B}_{\lambda}^{'}$ are obtained by $-\log\left|\dot{B}_{\lambda}^{'}\right|$.

As $A\tau \to \infty$, $\dot{B}_{\lambda}^{'}$ increases significantly (and symmetrically) for $\theta < 180^{o}$ and decreases significantly (and symmetrically) for $\theta > 180^{o}$. For substantial angular departures from 180º, the time rate of change of spectral radiance decreases symmetrically. The abrupt color change at $\theta$ = 180º in Figure 11 suggests a discontinuity in $\dot{B}_{\lambda}^{'}$. However, there are no values of $A$, $\tau$ or $\theta$ for which the denominator of eq. (18) is zero. Thus, there is no discontinuity. The change is over a scale which cannot be resolved on the graph. As expected, when $A\tau = 0$, the non-inertial relativistic spectral radiance is constant.

## 4. The Relativistic Wien's Displacement Law (RWDL)

The derivation of the Relativistic Wien's Displacement Law employs the same methodology as the stationary case for both the inertial and non-inertial reference frames, and can be easily derived from the relativistic spectral radiance in wavelength space, eq. (10).

### 4.1 RWDL in Inertial Reference Frames

Letting $Q = \frac{hc}{k_B}(\beta_t - \beta_z \cos\theta)$ and setting the derivative (with respect to $\lambda$) of the relativistic spectral radiance in wavelength space, eq. (10), equal to zero yields:

$$\frac{dB'_{\lambda}}{d\lambda} = \frac{Q\exp\left(\frac{Q}{\lambda}\right)}{\exp\left(\frac{Q}{\lambda}\right)-1} - 5 = 0 \tag{20}$$

This is the familiar transcendental equation that arises in the Wien's Displacement Law derivation for stationary observers. Therefore:

$$b = \frac{Q}{\lambda'_{\max}} \approx 4.965224231....$$

and

$$\lambda'_{\max} = \frac{hc}{bk_B}\left(\beta_t - \beta_z \cos\theta\right) \tag{21}$$

Rewriting eq. (21), in terms of *V*:

$$\lambda'_{\max} = \frac{hc}{bk_B T_o \sqrt{1-V^2}}\left(1 - V\cos\theta\right) = D^{-1}\lambda_{\max} \tag{22}$$

where $\lambda_{\max}$ is the wavelength of maximum irradiance in the radiation frame.

(21) and (22) are the Relativistic Wien's Displacement Law.

In the non-relativistic limit $(V << 1)$:

$$\lambda'_{\max} \sim \frac{hc}{bk_B T_o} - \frac{hc\cos\theta}{bk_B T_o}V + O\left(V^2\right) \tag{23}$$

where the first term in eq. (23) is $\lambda_{\max}$ (the radiation frame Stefan-Boltzmann Law).

Also, as $V \to 1$, $\lambda'_{\max} \to 0$ when $\theta = 0$, otherwise $\lambda'_{\max} \to \infty$. Consequently, when the velocity vector and the incident radiation vector are parallel (i.e. the observer is approaching the source directly), the frequency of radiation is completely blue shifted. Thus, the wavelength of maximum irradiance radiation in the observer frame approaches zero, and the frequency becomes infinite. However, for all off-axis observations of a moving radiation source, the wavelength becomes infinite and the frequency approaches zero because for increasing $\theta$, the redshifted component of the velocity vector dominates for increasing *V*. Since the RWDL is independent of temperature inflation and depends exclusively on Doppler shifting, it obviously exhibits the same behavior as pure Doppler shifting. This result is shown in Figure 12.

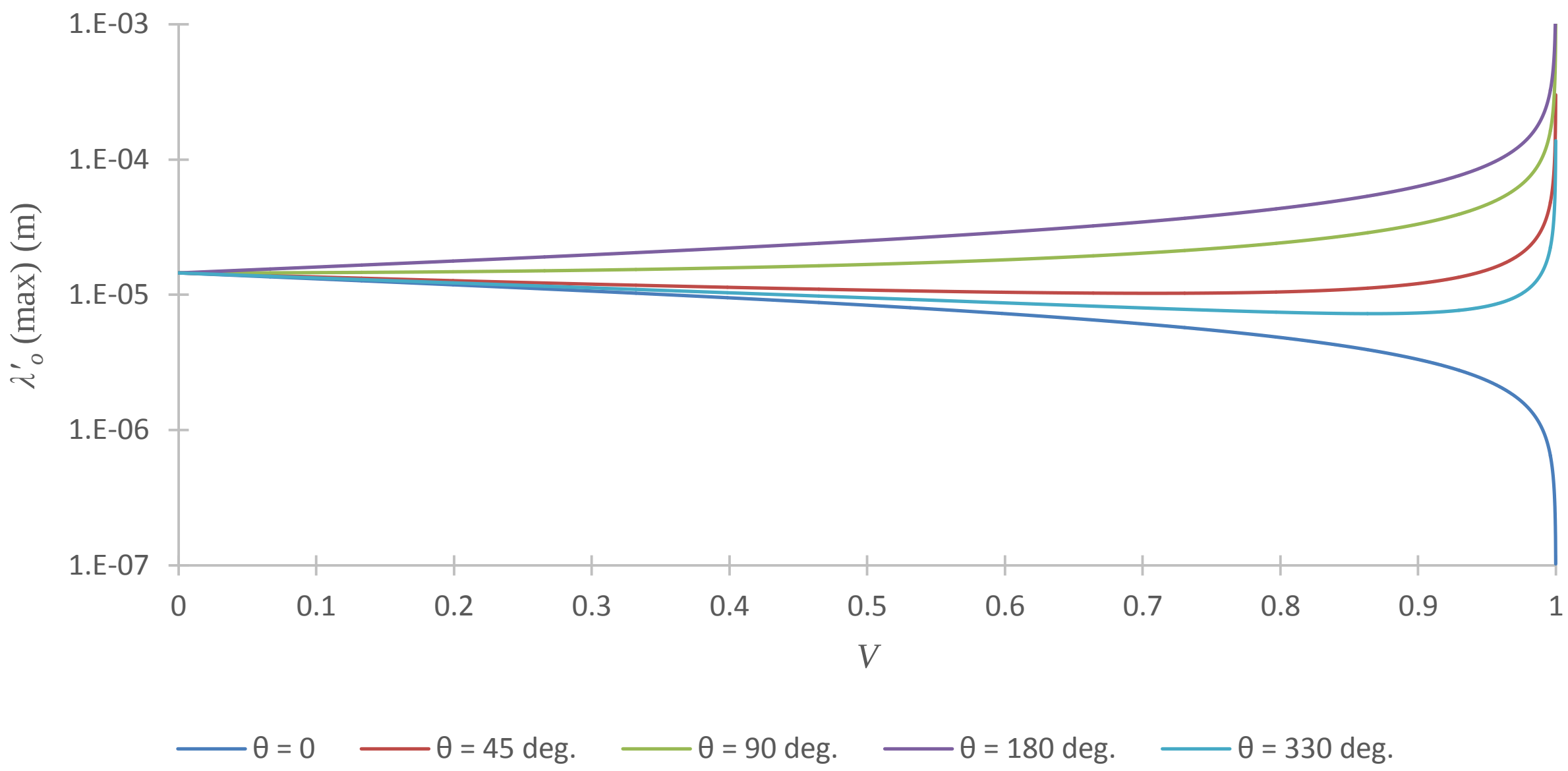

Figure 12: Wavelength of maximum irradiance versus speed for a radiation source with a proper temperature of 5000 K. The wavelength of maximum irradiance for the source when $V = 0$ is 0.5796 $\mu$m.

The angle which results in the maximum wavelength is expectedly the pure redshift case $\left(\theta = 180^o\right)$ and occurs from:

$$\frac{d\lambda'_{\max}}{d\theta} = \frac{V\sin\theta}{\sqrt{1-V^2}} = 0 \tag{24}$$

and thus,

$$\lambda'_{\max}\left(\theta = 180^o\right) = \frac{hc}{bk_BT_o}\sqrt{\frac{1+V}{1-V}} \tag{25}$$

as required.

### 4.2 RWDL in Non-Inertial Reference Frames

When the Relativistic Wien's Displacement Law is considered in a non-inertial reference frame, eqs. (17) and (22) are combined to give:

$$\lambda'_{\max} = \frac{hc}{bk_BT_o}\cosh(A\tau)(1-\tanh(A\tau)\cos\theta) = \cosh(A\tau)(1-\tanh(A\tau)\cos\theta)\lambda_{\max} \tag{26}$$

The angular dependence of the non-inertial, but constant-direction, RWDL is a simple cosine function, and its dependence on constant-direction acceleration and proper time is ultimately the same as the velocity dependence in eq. (22).

The time rate of change of the wavelength of maximum irradiance with a constant acceleration magnitude and a constant time rate of change of direction is:

$$\dot{\lambda}'_{\max} \equiv \frac{d\lambda'_{\max}}{\mathrm{d}\tau} = \frac{hc}{bk_B T_o}\left[A\sinh(A\tau)\left(1-\tanh(A\tau)\cos(\dot{\theta}\tau)\right) + A\cosh(A\tau)\cos(\dot{\theta}\tau) + \dot{\theta}\sinh(A\tau)\sin(\dot{\theta}\tau)\right]^{\mathrm{v}} \quad (27)$$

Clearly, if both the speed and direction components of the acceleration 4-vector are zero, then $\dot{\lambda}'_{\max} = 0$. Plots of eq. (27) are shown in Figure 13, Figure 14 and Figure 15.

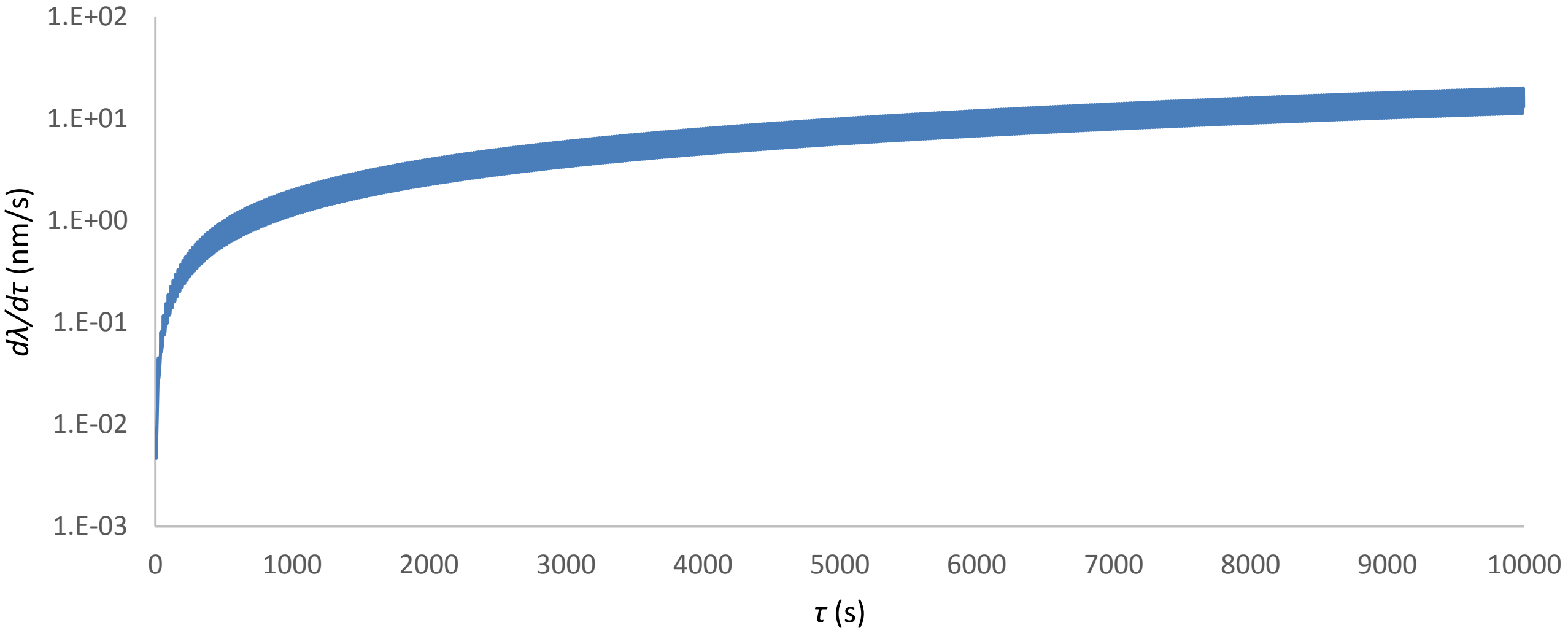

Figure 13: Time rate of change of wavelength of maximum irradiance versus proper time using an iterative scheme of 10,000 timesteps ($d\tau$ = 1s). $T_o$ = 5000 K, $A = 10^{-5}$, $\dot{\theta} = \frac{\pi}{9} \cdot \mathrm{s}^{-1}$.

[v] Note: $\theta = \dot{\theta}\tau$

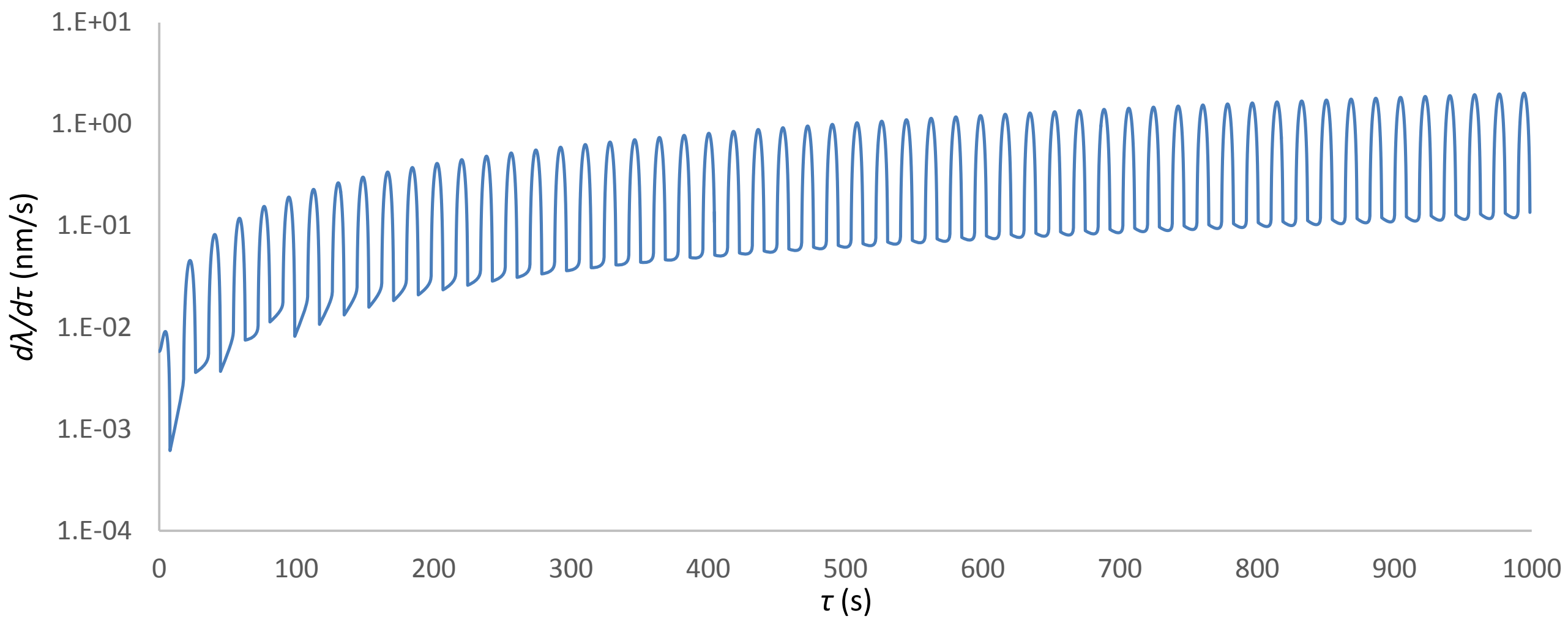

Figure 14: Time rate of change of wavelength of maximum irradiance versus proper time using an iterative scheme of 10,000 timesteps ($d\tau$ = 0.1s). $T_o$ = 5000 K, $A = 10^{-5}$, $\dot{\theta} = \frac{\pi}{9} \cdot \text{s}^{-1}$. (See discussion regarding the apparent jaggedness for τ.

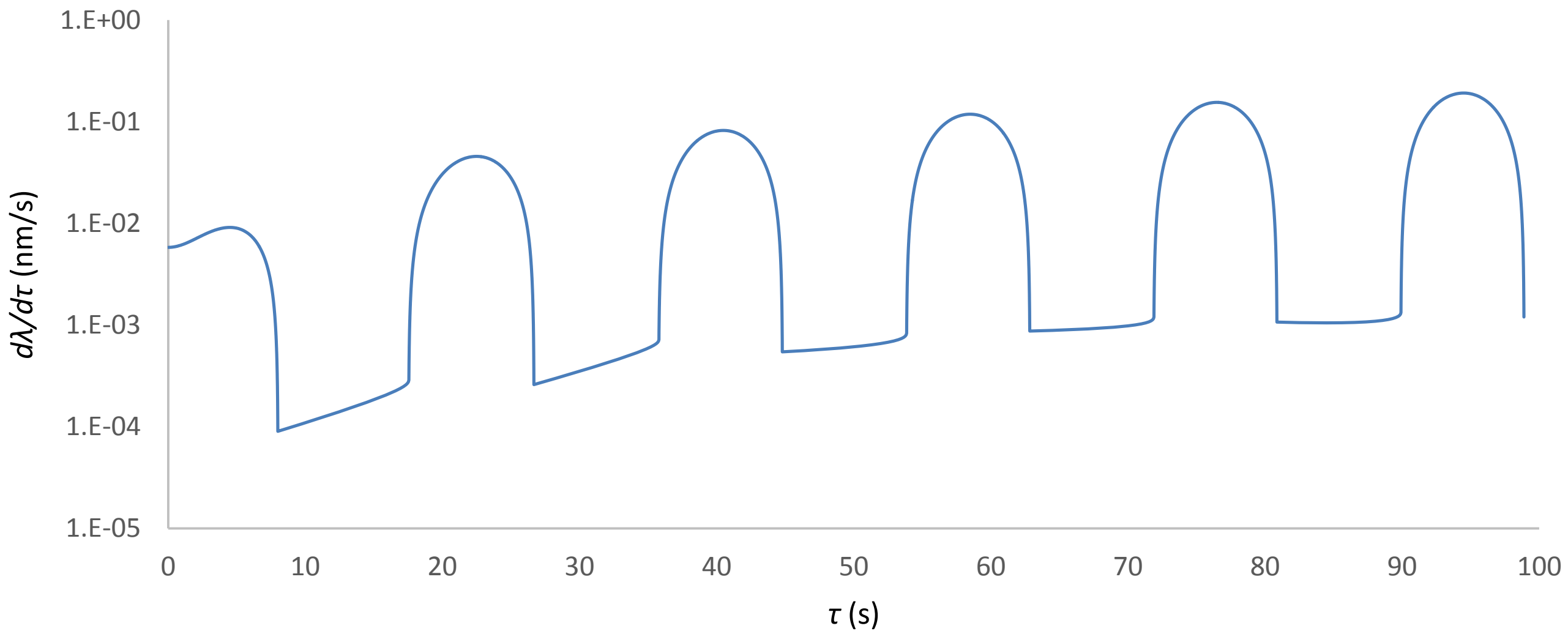

Figure 15: Time rate of change of wavelength of maximum irradiance versus proper time using an iterative scheme of 10,000 time steps ($d\tau$ = 0.01s). $T_o$ = 5000 K, $A = 10^{-5}$, $\dot{\theta} = \frac{\pi}{9} \cdot \text{s}^{-1}$.

The non-trivial periodicity of $\dot{\lambda}'_{\max}$ produces a function with an average period, in the above-mentioned example, of 0.06 s. The numeric noise which is evident at a resolution of 0.1 s is largely smoothed out at a resolution of 0.01 s.

## 5. The Relativistic Stefan-Boltzmann Law

Attempts to derive the Relativistic Stefan-Boltzmann Law have not relied on inverse temperature. Veitsman [3] did not rely on the invocation of the inverse temperature 4-vector, and asserts the necessity of accounting for the tensor character of temperature. However, his specific solution for $\theta = 0$ and his general solution for all $\theta$ do not agree, and numeric differences due to renormalization from using different coordinate systems is not, in contrast to his claim, a viable explanation for this discrepancy.

### 5.1 Inertial Reference Frames

The derivation of the Relativistic Stefan-Boltzmann Law can be accomplished by considering the power $(P')$, the area $(A')$, and the irradiance $(S')$ in frequency space, in a matter analogous to the derivation of the stationary Stefan-Boltzmann Law.

$$S' = \frac{P'}{A'} = \int_0^{\infty}\int B'_{\nu} d\Omega' d\nu' \tag{28}$$

From eqs. (9) and (28) [vi]:

$$S' = \int_0^{\infty}\int_0^{\frac{\pi}{2}} \frac{\frac{2h\nu^3}{c^2}}{\exp\left[\frac{h\nu}{k_B}(\beta_t - \beta_z\cos\theta)\right]-1} \left[\gamma(1 - V\cos\theta)\right]^{-3}\cos\theta\sin\theta d\theta d\nu \int_0^{2\pi} d\phi \tag{29}$$

By setting $\alpha = \frac{2h}{c^2}\left[\gamma(1-V\cos\theta)\right]^{-3}\cos\theta\sin\theta\int_0^{2\pi} d\phi = \frac{4\pi h}{c^2}\left[\gamma(1-V\cos\theta)\right]^{-3}\cos\theta\sin\theta$, letting $R = \beta_t - \beta_z\cos\theta$, and setting $x = \frac{h}{k_B}R\nu$, eq. (29) becomes:

$$S' = \int_0^{\infty}\int_0^{\frac{\pi}{2}} \frac{\alpha\nu^3}{e^x - 1} d\theta d\nu \tag{30}$$

$$S' = \frac{k_B^4}{h^4}\int_0^{\infty}\int_0^{\frac{\pi}{2}} \frac{\alpha x^3}{R^4(e^x - 1)} d\theta dx \tag{31}$$

---

[vi] The integral over $\phi$ is independent of $V$ because the motion is azimuthally constant. The $\cos\theta\sin\theta$ product arises from Lambert's Cosine Law ($\cos\theta$) and solid angle integration ($\sin\theta$).

The integral over *x* yields $\zeta(4)=\frac{\pi^4}{15}$, where $\zeta$ is the Riemann zeta function.

Thus:

$$S' = \frac{\pi^4 k_B^4}{15h^4}\int_0^{\frac{\pi}{2}} \frac{\alpha}{R^4}\,d\theta \qquad (32)$$

which becomes:

$$S' = \frac{4\pi^5 k_B^4}{15c^2h^3}\int_0^{\frac{\pi}{2}} \frac{\left[\gamma(1-V\cos\theta)\right]^{-3}\cos\theta\sin\theta}{(\beta_t-\beta_z\cos\theta)^4}\,d\theta \qquad (33)$$

From eqs. (7), (8), and expanding eq. (33):

$$S' = \frac{2\pi^5 k_B^4}{225c^2h^3}\left[\frac{(1+V)^2(1-V^2)^{\frac{3}{2}}(V^4-6V^3+15V^2-20V+15)}{(1-V)^4}\right]T_o^4 \qquad (34)$$

Eq. (34) is the inertial Relativistic Stefan-Boltzmann Law, which can also be rewritten in terms of the rest frame irradiance, *S*.

$$S' = \frac{S}{15}\left[\frac{(1+V)^2(1-V^2)^{\frac{3}{2}}(V^4-6V^3+15V^2-20V+15)}{(1-V)^4}\right] \qquad (35)$$

In the non-relativistic limit:

$$S' \sim \frac{2\pi^5 k_B^4 T_o^4}{15c^2h^3}+\frac{28\pi^5 k_B^4 T_o^4}{45c^2h^3}V+O(V^2) \qquad (36)$$

where the first term in eq. (36) is expectedly the radiation frame Stefan-Boltzmann Law.

Also, in eq. (34), when $V \to 1$, relativistic beaming and temperature inflation cause $S' \to \infty$, also as expected. Illustrated in Figure 16 is the irradiance of an inertial relativistic radiation source.

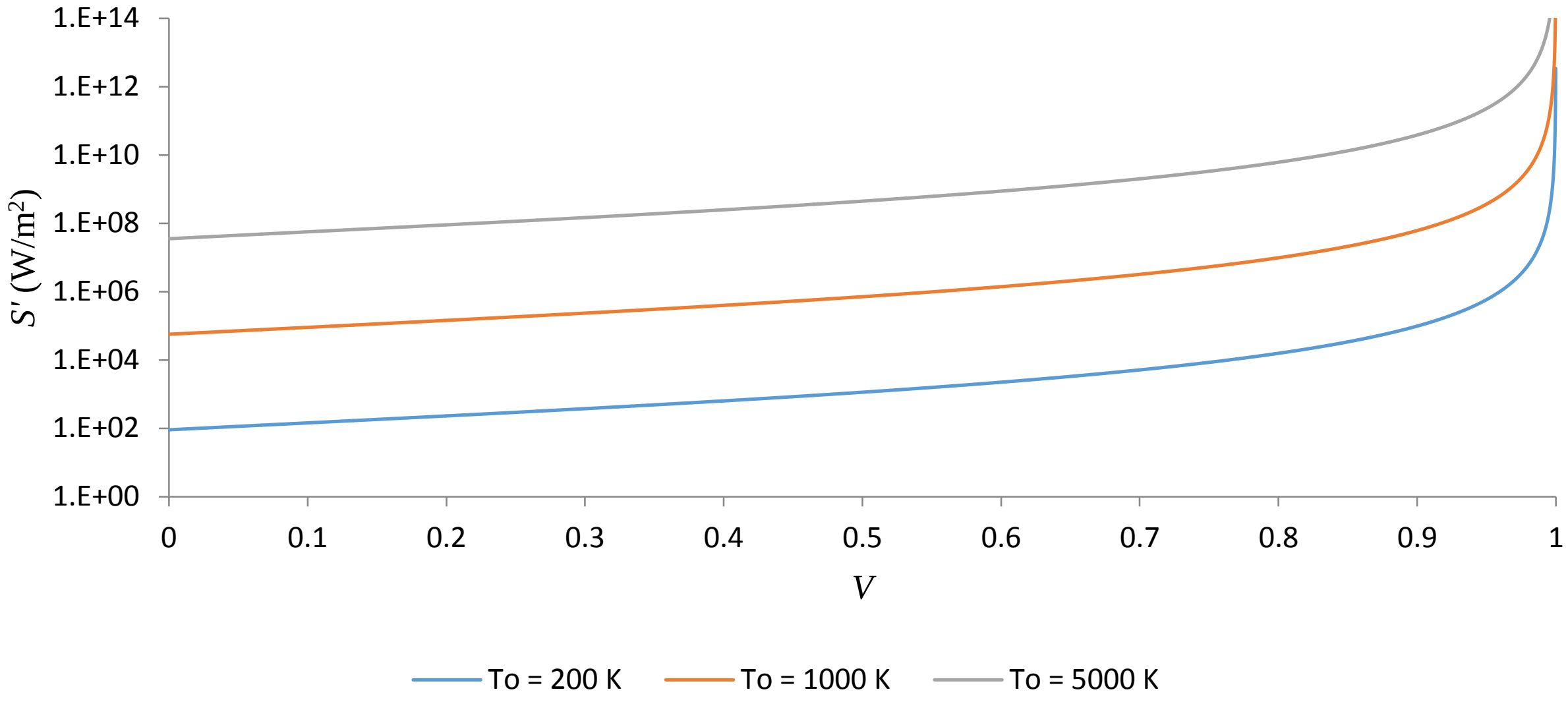

Figure 16: The inertial Relativistic Stefan-Boltzmann Law showing irradiance versus speed for a relativistic radiation source at three different temperatures. The $S'$-axis intercepts represent the irradiances in the radiation frame (90.3 W/m$^2$ for 200 K, $5.64\times10^5$ W/m$^2$ for 1000 K, $3.53\times10^7$ W/m$^2$ for 5000 K).

### 5.2 Non-Inertial Reference Frames

When the proper acceleration (with constant direction) is considered, eq. (37) results from eqs. (17) and (34), and is plotted in Figure 17.

$$S' = \frac{2\pi^5 k_B^4}{225c^2h^3}\left[\frac{\left(\begin{array}{l}\tanh^6(A\tau)-4\tanh^5(A\tau)+4\tanh^4(A\tau)+4\tanh^3(A\tau)\\-10\tanh^2(A\tau)+10\tanh(A\tau)+15\end{array}\right)(1+\tanh(A\tau))}{\cosh(A\tau)(1-\tanh(A\tau))^3}\right]T_o^4 \qquad (37)$$

As expected, when $A\tau = 0$, $S' = S$, and when $A\tau \to \infty$, $S' \to \infty$.

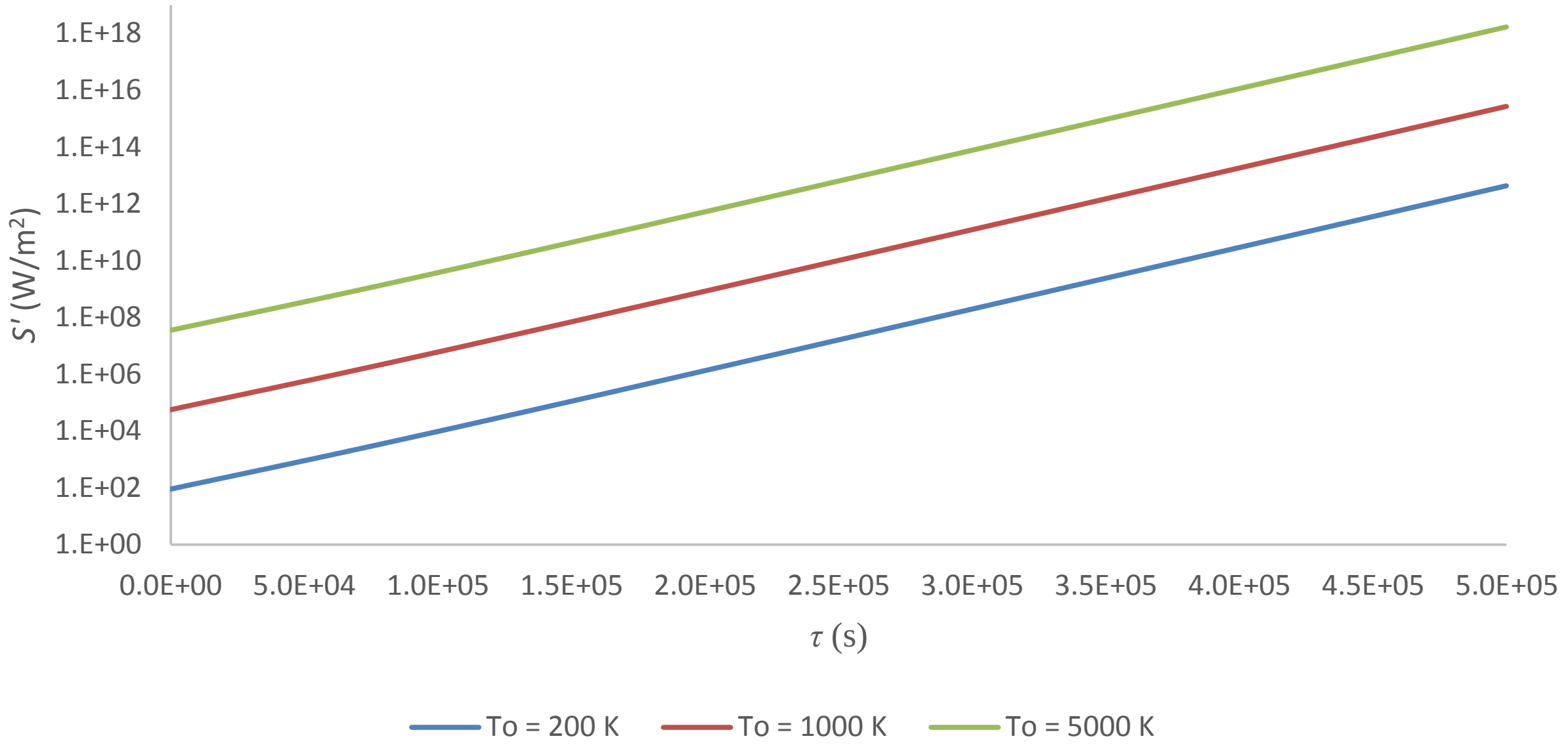

Figure 17: The inertial Relativistic Stefan-Boltzmann Law showing irradiance versus time for a non-inertial relativistic radiation source at three different temperatures. The *S'*-axis intercepts represent the irradiances in the radiation frame. $A = 10^{-5}$.

The time rate of change of the Relativistic Stefan-Boltzmann Law requires the proper time derivative of eq. (37) $\left(\dot{S}' \equiv \frac{dS'}{d\tau}\right)$, which produces the cumbersome result:

$$\dot{S}' = \frac{2\pi^5 k_B^4 T_o^4 A(\tanh(A\tau)+1)}{225c^2h^3\cosh(A\tau)(\tanh(A\tau)-1)^3} \times \left[5\tanh^7(A\tau) - 20\tanh^6(A\tau) + 22\tanh^5(A\tau) + 12\tanh^4(A\tau) - 42\tanh^3(A\tau) + 28\tanh^2(A\tau) - 35\tanh(A\tau) - 70\right] \quad (38)$$

When *A* = 0, the irradiance is constant. However, in the non-relativistic limit when $A \neq 0$ and $\tau \sim 0$:

$$\dot{S}' \sim \frac{28\pi^5 k_B^4 T_o^4}{45c^2h^3} A \quad (39)$$

Perhaps unexpectedly, it is reasonable that both $S'$ and $\dot{S}'$ lack angular dependence, even though the 4-acceleration is, in general, directionally dependent. This arises because the strict definition of $\dot{S}' \equiv \frac{dS'}{d\tau}$ requires the proper time differentiation of a function from which all angular dependence has been removed by the prior $d\Omega$ integration. Defining $\dot{S}' \equiv \int_0^\infty \int \dot{B}_\nu' \, d\Omega \, d\nu$ does not adhere to the required proper time derivative definition of $\dot{S}'$, in part because the order of the differentiation and integration must be preserved.

Figure 18 illustrates the proper time rate of change of the irradiance.

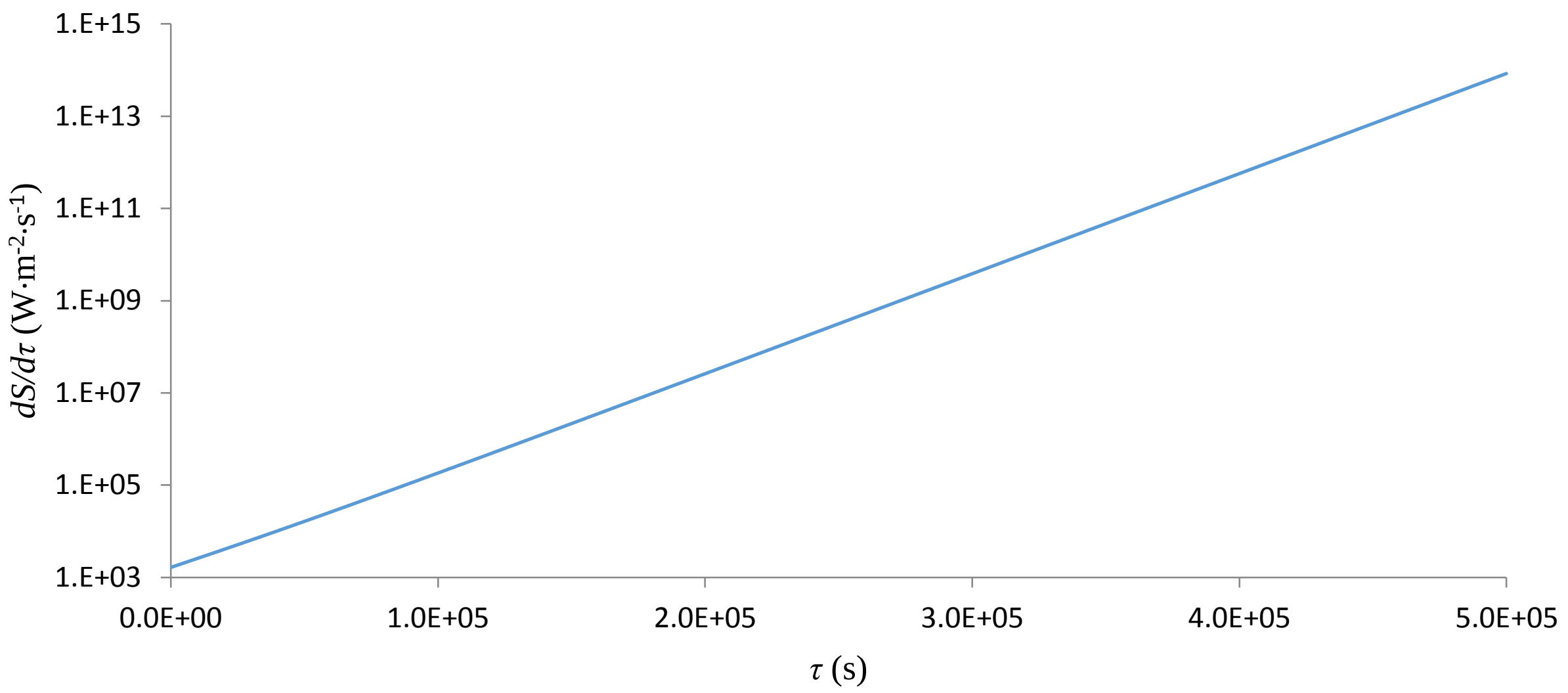

Figure 18: Proper time rate of change of irradiance of a relativistic blackbody. $T_o$ = 5000 K, $A = 10^{-5}$.

## 6. Observer and Radiation Frame Equivalences

The relativistic Planckian spectrum poses numerous questions of interest. Among them is whether there are non-trivial solutions for which the observer frame and radiation frame spectral radiances, wavelengths of maximum irradiance, and irradiances are equal.

### 6.1 Relativistic and Radiation Frame Spectral Radiances

The case of equal relativistic and radiation frame spectral radiances comes from equating eqs. (12) and (15) [vii]. Eq. (40) is the ratio of eqs. (12) and (15) set equal to 1.

$$\frac{\left[\exp\left(\frac{hc}{k_B \lambda T_o}\right)-1\right]\left(1-V^2\right)^{\frac{3}{2}}}{\left(1-V\cos\theta\right)^3\left\{\exp\left[\left(\frac{hc}{k_B \lambda T_o}\right)\frac{\left(1-V\cos\theta\right)}{\left(1-V^2\right)^{\frac{1}{2}}}\right]-1\right\}}=1 \tag{40}$$

Obviously, neither $V$ nor $\theta$ can be isolated as a closed-form function. However, a contour plot of eq. (40) is shown in Figure 19, and it is clear (as it was in Figure 2 and Figure 3) that there are non-zero speeds, as well as angles, for which $B'$ is equal to $B$, or even less than $B$.

[vii] This is in wavelength space. The case for frequency space is analogous.

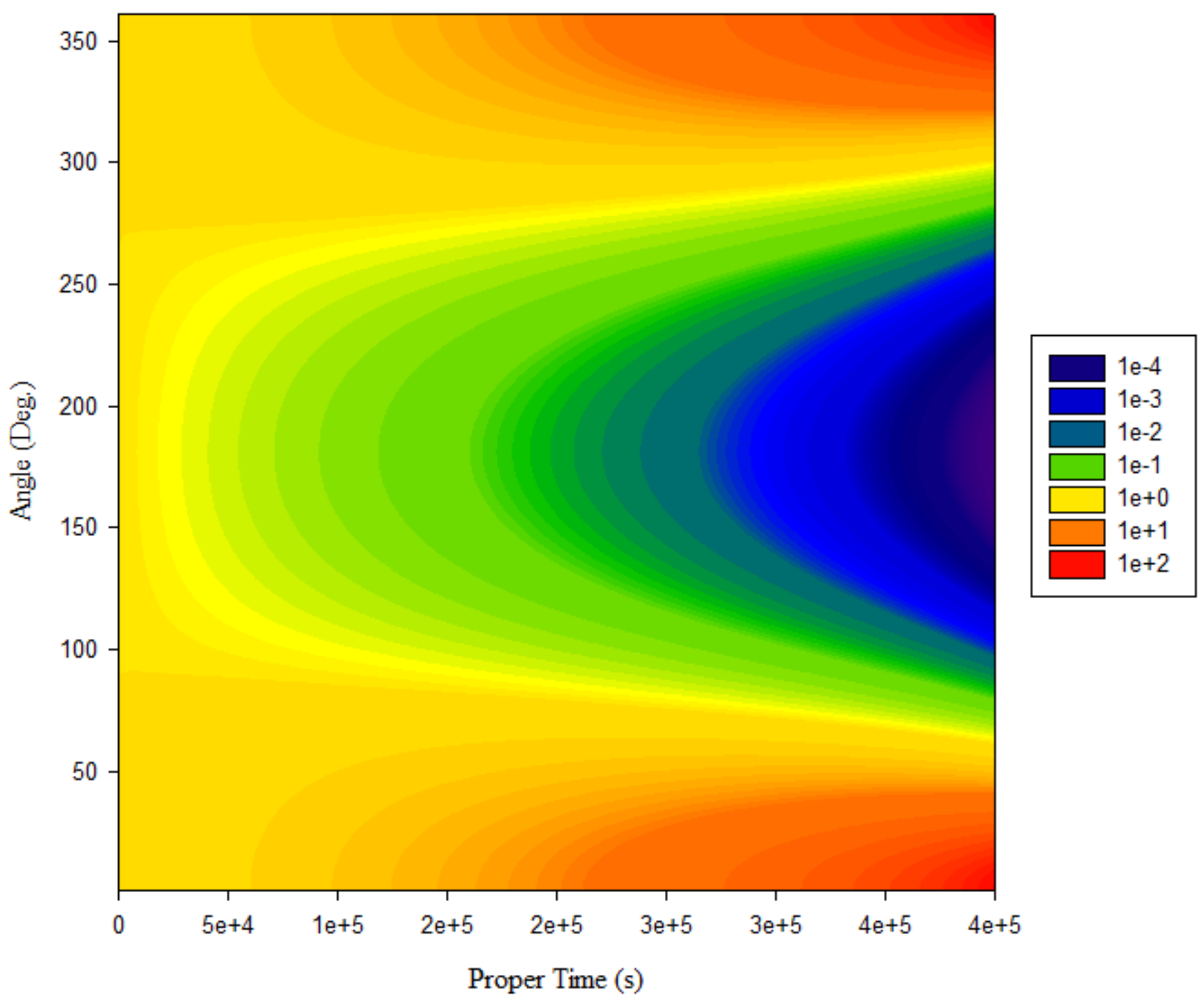

Figure 19: Contour plot of the ratio of inertial relativistic and radiation frame spectral radiances for a 5000 K blackbody observed at its wavelength of maximum irradiance (0.5796 $\mu$m). The yellow region indicates the $V$ and $\theta$ values at which the inertial relativistic spectral radiance is equal to the radiation frame spectral radiance. The green and blue regions represent the $V$ and $\theta$ values for which $B'$ is less than $B$. The region of smallest spectral radiance is actually $B'/B \leq 10^{-4}$.

### 6.2 Wavelengths of Maximum Irradiance

Far simpler is determining $V$ and $\theta$ for which the wavelength of maximum irradiance in the observer frame is equal to the wavelength of maximum irradiance in the radiation frame

From eq. (22):

$$\cos\theta = \sqrt{\frac{1+V}{1-V}} \tag{41}$$

which has no subluminal solution.

When eq. (22) is solved for $V$:

$$V = \frac{\cos^2\theta - 1}{\cos^2\theta + 1} \tag{42}$$

However, since $0 \leq V < 1$, $V = 0$ is the only physical solution to eq. (42). Therefore, as expected, there is no frame which isn't comoving with the radiation frame in which $\lambda'_{max} = \lambda_{max}$ .

### 6.3 Irradiances

The solution for the irradiance from the Relativistic Stefan-Boltzmann Law comes from eq. (35). The only real solution to eq. (43) (plotted in Figure 20) is $V = 0$. Expectedly, there is no frame which isn't comoving with the radiation frame in which the irradiance equals the irradiance in the radiation frame.

$$S^{'} = \frac{S}{15}\left[\frac{(1+V)^{2}(1-V^{2})^{\frac{3}{2}}(V^{4}-6V^{3}+15V^{2}-20V+15)}{(1-V)^{4}}\right] = S \tag{43}$$

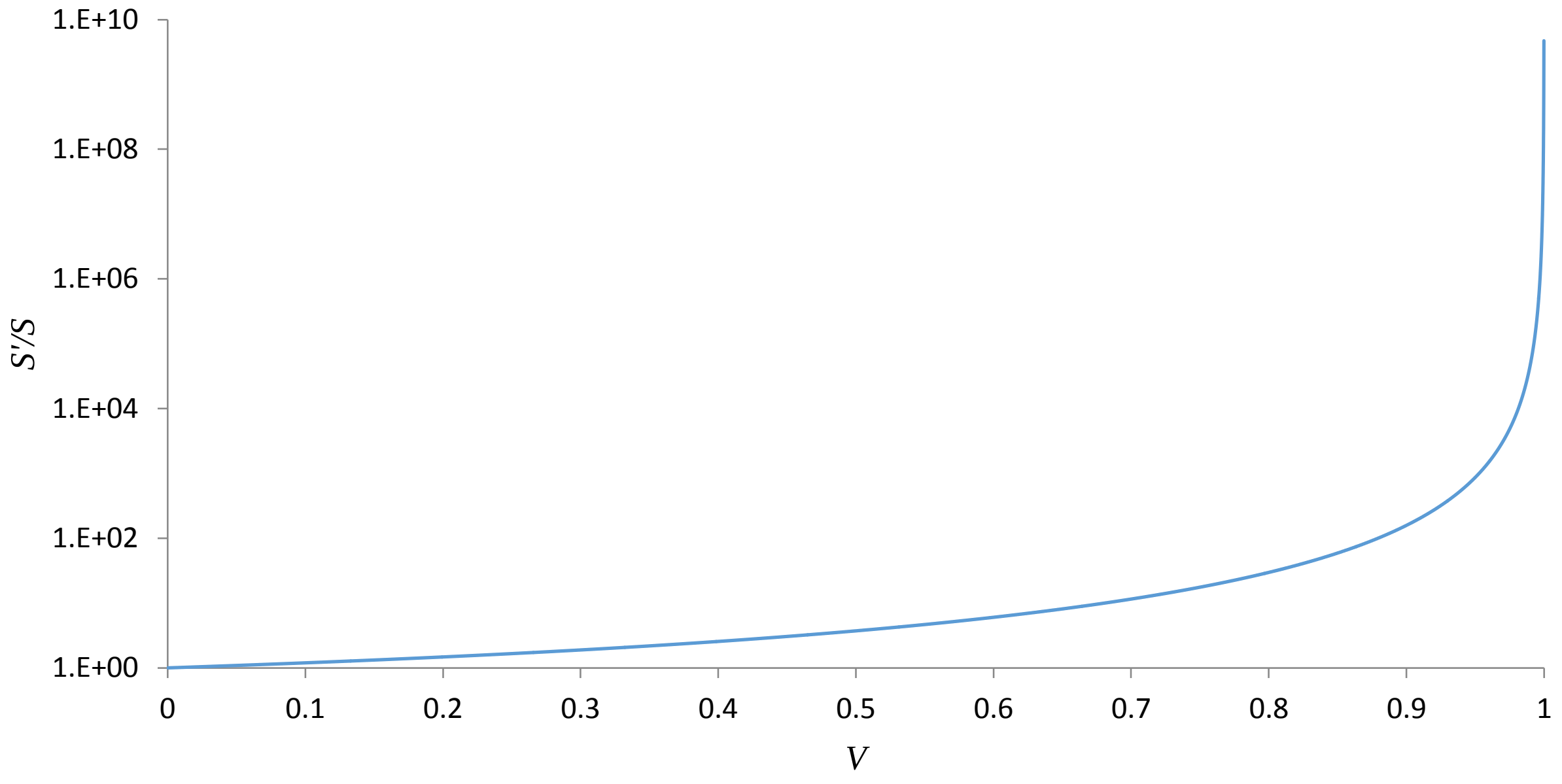

Figure 20: Ratio of irradiances in the observer and radiation frames as a function of speed for a blackbody. Clearly, only for $V >> 0$ does the departure from unity of $S'/S$ become significant.

## Conclusions

The Relativistic Planck's Law, the Relativistic Wien's Displacement Law, and the Relativistic Stefan-Boltzmann Law have been established in inertial and non-inertial reference frames by invoking the inverse temperature 4-vector, 4-acceleration, relativistic beaming, Doppler shifting, and, when required, the appropriate proper time derivatives. In the low velocity limit of the relativistic blackbody spectrum, the corresponding and well-established stationary blackbody spectrum has been shown to emerge for each of the aforementioned relativistic laws. The Relativistic Wien's Displacement Law was shown to be independent of temperature inflation and entirely dependent on Doppler shifting. In each case, the high velocity limit of the relativistic blackbody spectrum produced the expected zero or infinite outcome.

The angular periodicity of the Relativistic Planck's Law was determined, and further work needs to be done to elucidate the emergent picture. The relativistic versions of Planck's Law, Wien's Displacement Law, and the Stefan-Boltzmann Law were compared to the stationary versions, and it was determined that only in the case of spectral radiance are there non-trivial solutions by which the descriptions produce equal results.